 \documentclass[final,3p,times]{elsarticle}

\usepackage{amssymb}
\usepackage{lipsum}
\usepackage{multirow}
\usepackage{graphicx}
\usepackage{amssymb}
\usepackage{amsmath}
\DeclareMathOperator{\sech}{sech}
\usepackage{amsthm}
\usepackage{natbib}
\usepackage[colorlinks=true,linkcolor=black, citecolor=blue, urlcolor=blue]{hyperref}
\biboptions{sort&compress}
\usepackage{subfig}
\usepackage{graphicx}
\usepackage{multirow}
\usepackage{longtable}
\usepackage{mathrsfs}
\DeclareMathAlphabet{\pazocal}{OMS}{zplm}{m}{n}

\newcommand{\Lb}{\pazocal{L}}

\journal{Annals of Physics}
\begin{document}

\begin{frontmatter}



\title{Quasinormal modes and greybody factors of charge black hole in bumblebee gravity model }
\author[first]{Yenshembam Priyobarta Singh}
\ead{priyoyensh@gmail.com}

  \author[first] {Jayasri Choudhury}
\ead{jayasrichou57@gmail.com}
          
\author[first]{Telem Ibungochouba Singh \corref{cor1}}
\ead{ibungochouba@rediffmail.com}
\cortext[cor1]{Corresponding Author}

\affiliation[first]{organization={Department of Mathematics, Manipur University},
            addressline={Canchipur}, 
            city={Imphal},
            postcode={795003}, 
            state={Manipur},
            country={Inida}}

\author[second,third,fourth]{ Dhruba Jyoti Gogoi }
\ead{moloydhruba@yahoo.in}
\affiliation[second]{Department of Physics, Moran College, Moranhat, Charaideo 785670, Assam, India.}
\affiliation[third]{Research Center of Astrophysics and Cosmology, Khazar University, Baku, AZ1096, 41 Mehseti Street, Azerbaijan.}
\affiliation[fourth]{Theoretical Physics Division, Centre for Atmospheric Studies, Dibrugarh University, Dibrugarh
786004, Assam, India.}

%
%
%
\begin{abstract}
In this paper, we investigate the Dirac field, scalar field and electromagnetic field perturbations of Reissner-Nordstorm-de Sitter (RNdS) and  Reissner-Nordstorm-(anti)-de Sitter (RNAdS)-like black holes within the frame work of Einstein-bumblebee gravity. The effective potential, greybody factor and quasinormal modes (QNMs) are also explored by using Dirac equation, Klein-Gordon equation and Maxwell's equation. We find that for RNdS-like black hole increasing the Lorentz violation parameter $L$ consistently leads to  decrease in the effective potential for all types of perturbations  but for RNAdS case the influence of $L$ varies depending on the types of perturbation. Further for both RNdS and RNAdS-like black holes, increasing charge $Q$  reduces the effective potential in all the perturbations. The greybody factors of all the types of perturbations are also discussed. The results show that the greybody factors depend on the shape of the effective potential: higher (lower) potentials gives lower (higher) greybody factors. 
 The QNMs frequencies of RNdS-like black hole for the massless field perturbations are discussed by using 6th order WKB approximation and Padé approximation.  We also analyze the time-domain profiles of the perturbations.  The effects of Lorentz violation parameter $L$ and charge $Q$  to the photon sphere radius and shadow radius are also discussed. It is noted that increasing $Q$ and $L$ reduce the rise of shadow radius for RNdS-like black hole.
 
\end{abstract}

\end{frontmatter}

\section{Introduction}

The general relativity has explained the gravitational phenomena in the area of the classical physics which has also coped with rigorous theoretical and experimental validation. The standard model of particle physics described the other three fundamental interactions on the quantum front. A comprehensive understanding of the natural world can be explained by these two theories. The unification of these two theories will be fundamental and its results lead us to a deeper understanding of nature. 
Refs. \cite{J.M Maldacena1998, S.S. Gubser1998,J.Alfaro2002,C.Rovelli1990} proposed the several quantum gravity theories but the experimental unification will be at Planck scale $(\sim 10 ^{19} Gev)$ to observe the quantum effect which is impossible to conduct such experiment. The Lorentz symmetry violation is useful for solving the problem of   irrenormalization of gravity theory. Using ether like vector $u^{\alpha}$, the Lorentz violation theory of Dirac particle in flat Euclidean space is investigated in \citep{Bluhm2015}. Since then many fruitful results of black hole in Lorentz violation theory have been derived in \cite{Y.L. Onika2022,Y.S.Priyobarta2022,N.Media2024,
Y.L.Onika2023,priyobarta2024b,N.Media2022,N.Media2025}.   The prototype of the bumblebee model which is a string-inspire framework frequency tensor induced spontaneous Lorentz symmetric breaking has been disused in \cite{V.A.Kostelecky1989}. The potential function $V (B^a B_a)$ acting on a vector field $B_a$ leads to the formation of spacetime Lorentz violation in the framework of Einstein-bumblebee gravity model \cite{V.A.Kostelecky2004}. 
 Casana et al. \cite{casana2018} presented the static black hole solutions in bumblebee gravity. Subsequently, various black hole solutions have been explored within the framework of the bumblebee gravity model in \cite{gullu2022,ding2022,ovgun2019,ding2021,jha2021,gogoi2022}.

Before 1974, people assumed that the black hole is an entity where the gravitational field is so strong, even light cannot escape from it. Hawking \cite{hawking1974} discovered that a black hole emits and creates particles near the event horizon when quantum gravity effect is taken into account. Since then,  this phenomenon is known as Hawking radiation. The gravitational potential generated by the black hole encounters this radiation which gives the reflections and transmissions of the Hawking radiation. Hence it makes the difference of the actual spectrum observed by an asymptotic observer from the black body spectrum. 
The quantity which describes the deviation of the Hawking radiation is known as the black hole greybody factor. Since then many researchers proposed the different methods for finding the greybody factors such as WKB approximation technique for high gravitational method \cite{M.K.Parikh2000,R.A.Konoplyaa2020}, matching method \cite{S.Fernando2005,W.Kim2008} and rigorous bound method \cite{M. Visser1999,P. Boonserm2008}.
Rigorous bound method will be applied in our paper.

It is also noted that the oscillation of black hole with complex frequency took place in the intermediate range due to non-radial perturbation. The oscillation is known as quasinormal modes (QNMs). It is also found that the several complex frequencies were produced by quasinormal ringing in which the real part of the frequency leads to the oscillation rate  and the magnitude of the imaginary part  leads to the damping rate.
The QNMs indicated the frequency of oscillations which heavily rely on the characteristic of  perturbed black hole like  mass, charge and spin  \cite{R.A.Konoplya2011,E.Berti2009}. It is well noted that all the perturbation equations of a family of black holes may be recast into a one-dimensional Schrödinger equation. The QNMs are the solutions of the Schrödinger wave equation with complex frequencies along with special boundary conditions which also represent purely ingoing near the horizon and completely outgoing at asymptotic  infinity. A unique information about the parameters of the black hole can be derived from the QNMs and it may also give a hint into the black holes of quantum nature \cite{O.Dreyer2003,M. Maggiore2008}. The QNMs may also be described in terms of the overtone $n$ and  multipole moment quantum number $l$. Many researchers have developed various methods for finding QNMs such as time domain method \cite{C.V.Vishveshwara1971,E.N.Dorband2006,davis1971}, direct integration in the frequency domain \cite{H.P. Nollert1992,N. Andersson1992}, continuous fractions method \cite{E.W. Leaver1985}, Pösch–Teller fitting method \cite{V. Ferrari1984}, Frobenius method \cite{R. A. Konoplya2011} and  WKB method \cite{B.F. Schutz1985, S. Iyer1987} etc. Ref. \cite{H.T. Cho2003} showed that WKB method gives more accurate results for both the real part and the imaginary part with $n \le \ell$.
Following the above methods, the QNMs of different kind of black holes are also discussed in \cite{Y.S.Priyobarta2024,S.Devi2020,K.J.Sohan2024,
K.J.Sohan2023,L.Jin2023,Al-Badawi2024,D.J. Gogoi2023,L.Jianhui2024,J-Z.Liu2025,A. Chowdhury2020,A. Uniyal2023,P.Boonserm2018,P.Boonserm2021}.

 Synge firstly studied \cite{J.L.Synge1966} the shadow cast by a Schwarzchild black hole. Ref. \cite{J.P.Lumine1979} also investigated the impact of the accretion disc on the shadow of the black hole while the shadow of Kerr black hole is discussed in \cite{J.M. Bardeen1973}. Since then, many interesting results have been developed in the study of black hole shadow \cite{A.de Vries1999,C.Bambi2012,F.Atamurotov2013,T.- T.Sui 2023} after taking the direct shadow images of $M87^*$ at the core of Vigro A galaxy and Sgr$A^*$ at the core of the milky way galaxy by using the event horizon Telescope \cite{K. Akiyama2019 1, K. Akiyama2019 2, K. Akiyama 2022}. Under the context of bumblebee gravity, black hole shadow is studied in \cite{Y.S.Priyobarta2024,A. Uniyal2023,maluf2021,izmailov2022,priyobarta2025}.

The QNMs and greybody factors of Schwarzschild black hole with cosmological constant under bumblebee gravity model have been studied  \cite{Y.S.Priyobarta2024, A. Uniyal2023}. However, the extension to charge black hole with  a cosmological constant, such as RNdS-like and RNAdS-like spacetimes have not been investigated before. This gap inspires our current work and the main motivation is to probe the effects of the Lorentz violation theory arising from  a nonzero vacuum expectation value of the bumblebee vector field  on fundamental physical observables such as  quasinormal modes, greybody factors and black hole shadow. Moreover, we intend to study the perturbations of scalar, electromagnetic and Dirac fields in  RNdS and RNAdS-like black hole within the frame work of Einstein bumblebee gravity model.  

The organization of this paper is as follows : In Section   \ref{sec 2}, the properties of RNdS and RNAdS-like black holes in bumblebee gravity model are discussed and derive the corresponding effective potentials. We investigate the perturbations of scalar and electromagnetic field and derive the effective potentials  of the black holes in Section \ref{sec 4}. Dirac field pertubation is analyze in Section \ref{sec dirac}. In Section  \ref{sec 5}, we calculated the greybody factors of RNdS-like black hole for the perturbations of scalar, electromagnetic and Dirac fields. The greybody factor for massless perturbation for different black holes are also discussed in Section  \ref{sec 6}. Applying WKB 6th order and Padé  approximation, we derive QNMs of RNdS-like black hole for all the perturbations in Section  \ref{sec 7}. In Section \ref{sec 8}, we investigate the time evolution profiles of the perturbations. The null geodesic and shadow radius for both RNdS and RNAdS-like black holes are also discussed  in Section  \ref{sec 9}. 
 Some conclusions are given in section  \ref{sec 11}.

\section{Charged black hole with cosmological constant in bumblebee gravity }\label{sec 2}
A solution of charged black hole with a cosmological constant within the framework of bumblebee gravity was proposed in  \cite{J-Z.Liu2025}.
The spacetime metric of Reissner-Nordstorm-like  black hole with a cosmological constant in bumblebee model is given by \cite{J-Z.Liu2025}
\begin{align} \label{eqn 1}
	ds^2= A(r) dt^2 -\dfrac{1+L}{A(r)} dr^2-r^2 (d\theta^2+\sin^2\theta d\phi^2),
\end{align}
where
\begin{align}\label{eqn 2}
A(r)=1-\dfrac{2M}{r}+\dfrac{2(1+L)}{2+L}\dfrac{Q^2}{r^2}-\dfrac{1}{3}(1+L)r^2\Lambda.
\end{align}
Here  $M,$ $Q$ and $\Lambda$ are the mass, charge and cosmological constant respectively. $L$ denotes the Lorentz violation parameter.  Eq.  \eqref{eqn 1} represents dS black hole or AdS black hole if $\Lambda> 0$  or $ \Lambda< 0$. In the limit $L\rightarrow 0$, \eqref{eqn 1} tends to the line element of original Reissner-Nordstrom-dS/AdS black hole.
The non vanishing covariant components of $g_{ab}$ are given by 
\begin{align}
&g_{00}=A(r) ,\hspace{1cm}
g_{11}=-\dfrac{1+L}{A(r)},\hspace{1cm}g_{22}=-r^2,\hspace{1cm}g_{33} = -r^2 \sin^2\theta.
\end{align}
 
 \begin{figure}[h!]
\centering
  \subfloat[\centering ]{{\includegraphics[width=170pt,height=170pt]{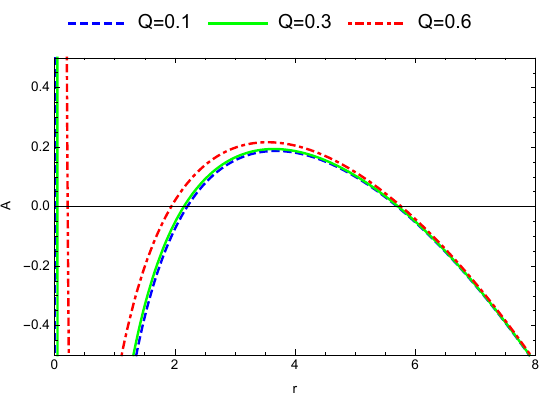}}\label{fig 1a}}
  \qquad
   \subfloat[\centering ]{{\includegraphics[width=170pt,height=170pt]{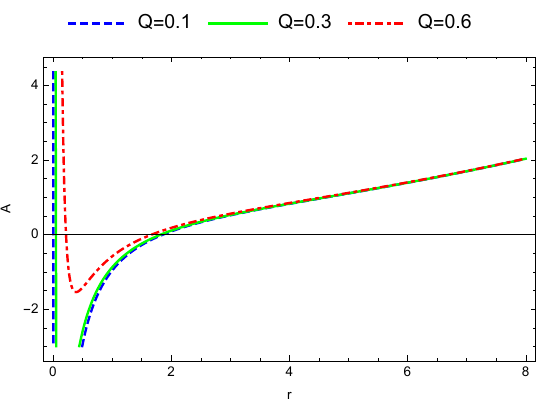}}\label{fig 1b}}
   \caption{Variation of metric function for different values of $Q$ with fixed (a) $M=1$, $\Lambda=0.05$ and $L=0.2$  (b) $M=1$, $\Lambda=-0.05 $ and $L=0.2$ .}
   \label{fig 1}
\end{figure}
\begin{figure}[h!]
\centering
  \subfloat[\centering ]{{\includegraphics[width=170pt,height=170pt]{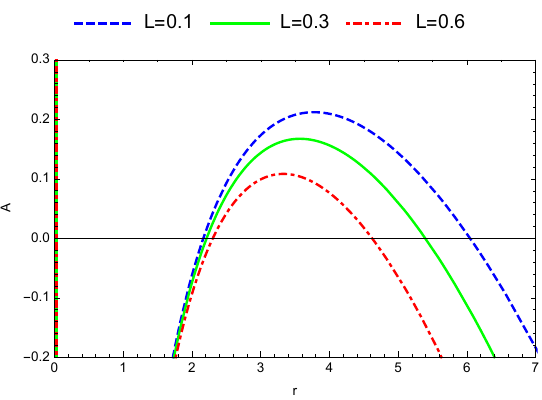}}\label{fig 2a}}
  \qquad
   \subfloat[\centering ]{{\includegraphics[width=170pt,height=170pt]{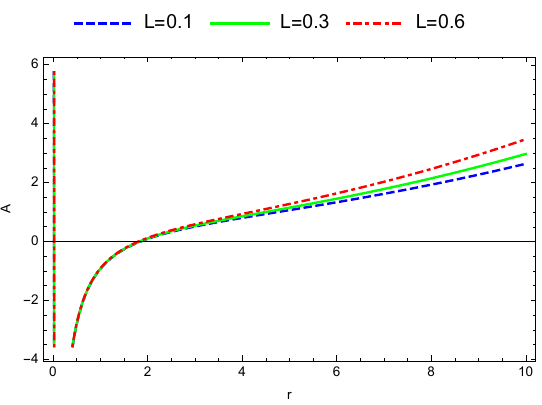}}\label{fig 2b}}
   \caption{Variation of metric function for different values of $L$ with fixed (a) $M=1$, $\Lambda=0.05$ and $Q=0.2$  (b) $M=1$, $\Lambda=-0.05 $ and $Q=0.2$ .}
   \label{fig 2}
\end{figure}
For RNdS-like black hole the metric function \eqref{eqn 2} can be expressed as 
\begin{align} 
&A(r)=\dfrac{-(1+L)\Lambda}{3r^2} (r-r_-) (r-r_+) (r-r_h) (r-r_c), ~~~~ 0<r_+<r_h<r_c<\infty \nonumber.
\end{align}
Here   $r_+$, $r_h$ and $r_c$ denote the three physical horizons:  Cauchy horizon, event horizon and cosmological horizon respectively. The unphysical horizon is $r_-$. For RNAdS-like black hole, there exist only two physical horizons: Cauchy horizon and event horizon. Thus the metric function of RNAdS-like black hole is expressed as
\begin{align}
&A(r)=\dfrac{-(1+L)\Lambda}{3r^2} (r-r_-) (r-r_{--}) (r-r_+) (r-r_h) ,~~~~ 0<r_+<r_h<\infty \nonumber.
\end{align}


Figs. \ref{fig 1} and \ref{fig 2} show the nature of metric function for both RNdS and RNAdS-like black holes for varying $Q$ and $L$ respectively. For RNdS-like black hole, the distance between event horizon and cosmological horizon increases with increasing $Q$, but it has opposite effect with increase in $L$. Further with increasing $Q$, the  distance between Cauchy horizon and event horizon decreases for RNAdS-like black hole.

\section{Scalar and electromagnetic field perturbation}\label{sec 4}
This section investigates the effective potentials of scalar and electromagnetic perturbations in RNdS and RNAdS-like black holes  within the framework of bumblebee gravity. To study the massive scalar field perturbation, we use the Klein-Gordon equation in curved space time which is defined by \cite{ibungochouba2015,peskin2018}
\begin{align}\label{eqn 23}
\dfrac{1}{\sqrt{-g}}\partial_{a}\left(\sqrt{-g}g^{ab}\partial
_b\psi \right)+m^2\psi=0.
\end{align}
For the electromagnetic field  perturbation,  the Maxwell equation is given by \cite{cardoso2001}
\begin{align}\label{eqn 24}
\dfrac{1}{\sqrt{-g}}\partial_{a}\left( F_{bc}g^{bd}g^{ca}\sqrt{-g}\right)=0,
\end{align}
where $F_{bc}$=$\partial_{b}A_c-\partial_{c}A_{b}$ and $A_a$ denotes the electromagnetic four-potential. 
After separation of variables and introducing the tortoise coordinate transformation defined as
   \begin{align}\label{eqn 14}
   d{r_*}=\dfrac{\sqrt{1+L}}{A}d{r},
\end{align}
 the radial parts of Eqs. (\ref{eqn 23}) and (\ref{eqn 24}) take the form of  Schrödinger-like equation as
\begin{align}\label{eqn Vse}
-\dfrac{d^2{\phi}}{d{r_*^2}}+V_{s,e}~ \phi=\omega^2\phi.
\end{align}
Here $\omega$ denotes the QNMs associated with the scalar and electromagnetic field $\phi$. The perturbations of QNMs are known as the ingoing wave for $r_*\rightarrow -\infty $,  $\phi\simeq e^{i\omega(t+r_*)}$ and the outgoing wave for  $r_*\rightarrow \infty $,  $\phi\simeq e^{i\omega(t-r_*)}$ \cite{konoplya2003}. The effective potential for the perturbation of the scalar field is obtained as
\begin{align}\label{eqn Vs}
V_s= A\left[\dfrac{1}{r(1+L)}\dfrac{d{A}}{d{r}}+\dfrac{l(1+l)}{r^2}+m^2 \right] 
\end{align}
and the effective potential for the perturbation of the electromagnetic field is also found as 
\begin{align}\label{eqn Ve}
V_e=\dfrac{l(1+l)}{r^2}A.
\end{align}
The two effective potentials derived from the Klein-Gordon equation and the electromagnetic field equation can be recast into a single equation as 
\begin{align}\label{eqn Vms}
V_{ms}=A\left[\dfrac{l(1+l)}{r^2}+(1-s^2)\left\lbrace  \dfrac{1}{r(1+L)}\dfrac{d{A}}{d{r}}+m^2\right\rbrace \right].
\end{align}
It is noted from the above equation that $s=0$ and $s=1$ give the effective potential associated with the scalar field perturbation and electromagnetic field perturbation respectively. $s$ and $l$ are the spin and angular momentum respectively.
\begin{figure}[h!]
\centering
  \includegraphics[width=170pt,height=170pt]{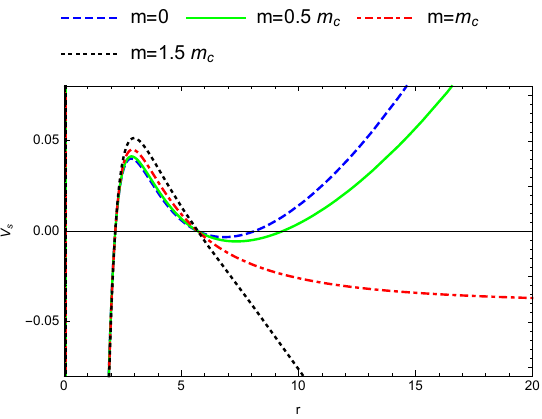}
   \caption{Variation of the effective potential for the massive scalar field for RNdS-like black hole for different values of the mass parameter of the field $m$. The physical parameters are chosen as  $M=1$, $\Lambda=0.05$, $L=0.2$,  $\ell=1$ and $Q=0.2$.}
   \label{fig SVM}
\end{figure}
 
\begin{figure}[h!]
\centering
  \subfloat[\centering ]{{\includegraphics[width=170pt,height=170pt]{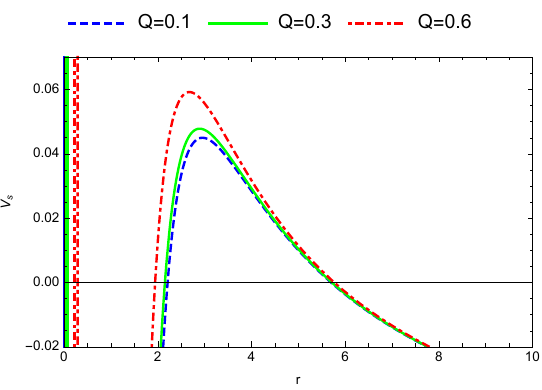}}\label{fig 5a}}
  \qquad
   \subfloat[\centering ]{{\includegraphics[width=170pt,height=170pt]{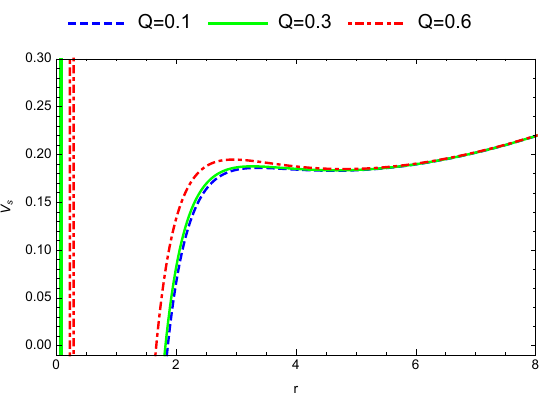}}\label{fig 5b}}
   \caption{Variation of the effective potential for the massive scalar field for different values of $Q$. The physical parameters are chosen as (a) $M=1$, $\Lambda=0.05$, $L=0.2$,  $\ell=1$ \text{and} $m=0.1$ (b) $M=1$, $\Lambda=-0.05$, $L=0.2$, $\ell=1$ \text{and} $m=0.1$.}
   \label{fig 5}
\end{figure}

\begin{figure}[h!]
\centering
  \subfloat[\centering ]{{\includegraphics[width=170pt,height=170pt]{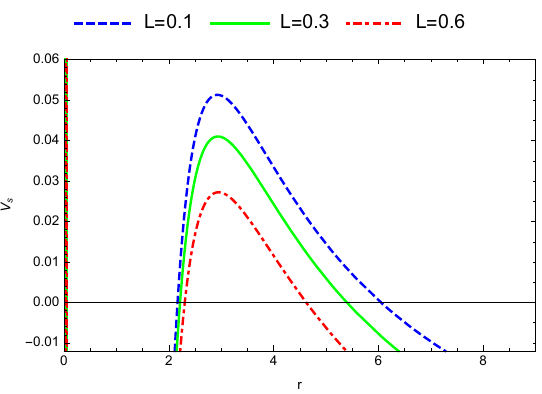}}\label{fig 6a}}
  \qquad
   \subfloat[\centering ]{{\includegraphics[width=170pt,height=170pt]{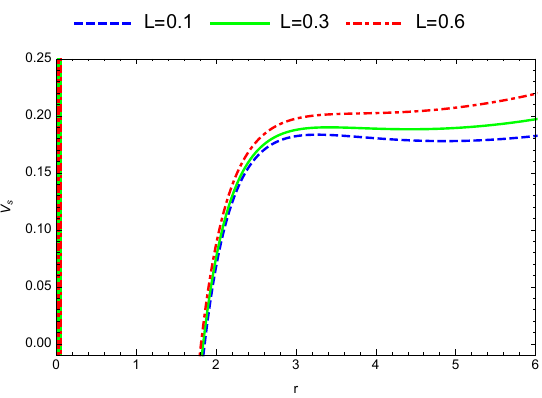}}\label{fig 6b}}
   \caption{Variation of the effective potential for the massive scalar field for different values of $L$. The physical parameters are chosen as (a) $M=1$, $\Lambda=0.05$, $Q=0.2$, $\ell=1$ \text{and} $m=0.1$  (b) $M=1$, $\Lambda=-0.05$, $Q=0.2$, $\ell=1$ \text{and} $m=0.1$.}
   \label{fig 6}
\end{figure}

\begin{figure}[h!]
\centering
  \subfloat[\centering ]{{\includegraphics[width=170pt,height=170pt]{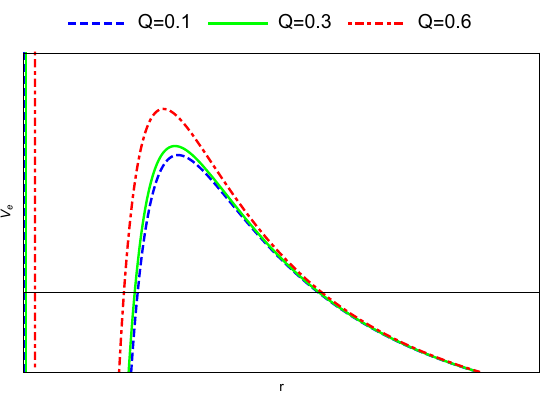}}\label{fig 7a}}
  \qquad
   \subfloat[\centering ]{{\includegraphics[width=170pt,height=170pt]{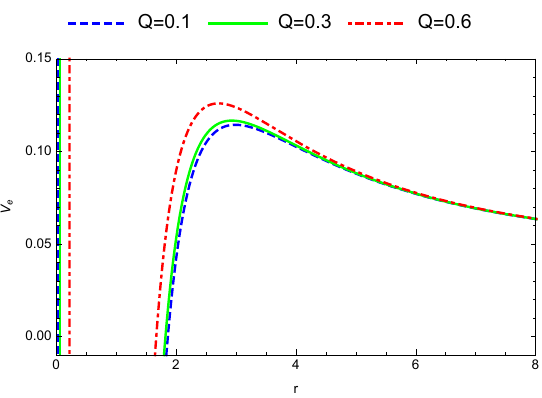}}\label{fig 7b}}
   \caption{Variation of the effective potential for the electromagnetic field for different values of $Q$. The physical parameters are chosen as (a) $M=1$, $\Lambda=0.05$, $\ell=1$ \text{and} $L=0.2$  (b) $M=1$, $\Lambda=-0.05$, $\ell=1$ \text{and} $L=0.2$.}
   \label{fig 7}
\end{figure}

\begin{figure}[h!]
\centering
  \subfloat[\centering ]{{\includegraphics[width=170pt,height=170pt]{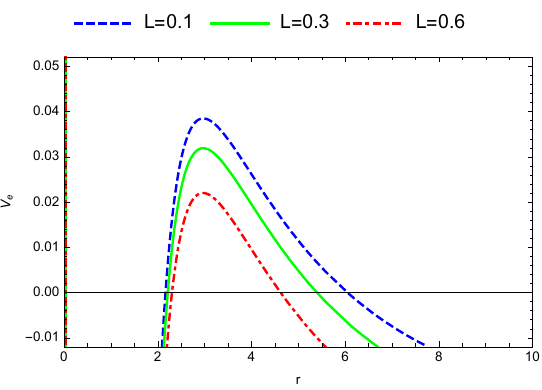}}\label{fig 8a}}
  \qquad
   \subfloat[\centering ]{{\includegraphics[width=170pt,height=170pt]{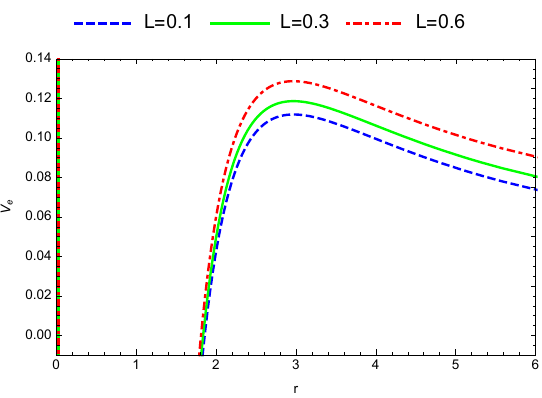}}\label{fig 8b}}
   \caption{Variation of the effective potential for the electromagnetic field for different values of $L$. The physical parameters are chosen as (a) $M=1$, $\Lambda=0.05$, $\ell=1$ \text{and} $Q=0.2$   (b) $M=1$, $\Lambda=-0.05$, $\ell=1$ \text{and} $Q=0.2$.}
   \label{fig 8}
\end{figure}

 In Fig. \ref{fig SVM}, we plot the effective potential of  scalar field for RNdS-like black hole for different values of $m$. It is observed that there exists a critical mass $m_c$, which marks a transition point in the behavior of the effective potential governing the dynamics of the scalar field. For massless and $m<m_c$, the effective potential  vanishes at  $r_+$, $r_h$, $r_c$ and another additional zero point $r_a$. Here $r_a$ is the point at which the second factor of $V_s$ vanishes. Further if $m>m_c$, there is no $r_a$ and the effective potential exhibits a single peak and it is monotonically decreasing for all $r>r_c$. Henceforth, we restrict our analysis to the regime 
$m>m_c$. 
To find the critical value of the mass parameter $m_c$ of the scalar particle, first we study the nature of the effective potential at spatial infinity. For this purpose, we focus on the leading behavior of the effective potential when $r \rightarrow \infty$ as
\begin{align}
V_s \sim \frac{1}{3}(1 + L) r^2 \Lambda \left( \frac{2}{3} \Lambda -m^2\right).
\end{align}
The critical value of the mass $m_c$ is the value of $m$ when the effective potential flattens. Thus, the critical mass $m_c$ is obtained as
\begin{align}\label{eqn mc}
m_{c} =\sqrt{ \frac{2}{3} \Lambda}.
\end{align}
From Eq. \eqref{eqn mc}, we know that the critical mass $m_c$ depends only on $\Lambda$.
Figs. \ref{fig 5}, \ref{fig 6} and \ref{fig 7}, \ref{fig 8} show the difference of effective potentials of massive scalar field and electromagnetic field perturbations for varying $Q$ and $L$ respectively.
Increasing the values of $Q$, the peak of effective potentials derived from massive scalar field perturbation and electromagnetic field perturbation increase for both RNdS and RNAdS-like black holes. Thus increasing $Q$ makes the radiation   more difficult   to escape, from the potential barrier, thereby  reducing the greybody factor. It is also shown that increasing the parameter $L$, the peak of effective potentials derived from both massive scalar and electromagnetic field perturbations decreases for RNdS black hole only, but it has an opposite effect for RNAdS-like black hole for both massive scalar and electromagnetic field  perturbations. The different effects of $L$ in RNdS and RNAdS-like black hole underscore the significant influence of $\Lambda$ on black hole dynamics within bumblebee gravity. For dS case, increasing $L$ facilitates the escape of radiation indicating the increase of the greybody factor but  the opposite occurs in AdS case.  

\section{Dirac field perturbation}\label{sec dirac}
The study of Dirac field perturbation in RNdS and RNAdS-like black holes under bumblebee gravity  provides valuable insights into how the  Lorentz symmetry violation influence the behavior of  spin-1/2 particles, like electrons and neutrinos, under extreme gravitational conditions. 
 To study the Dirac field perturbation we use
the Dirac equation with mass $m$ in a general background spacetime which is defined as  
 \begin{align}\label{eqn 4}
	\left[\gamma^a e^\alpha_a(\partial_\alpha+\Gamma_\alpha)-m\right]\Psi=0,
 \end{align}
where $\gamma^a$ are the gamma matrices, which are defined by
\begin{align}\label{eqn 5}
	\gamma^0=\begin{pmatrix}
	   -i & 0\\
	   	0 & i\\
	   \end{pmatrix}, \hspace{2cm}
	\gamma^j=\begin{pmatrix}
		0 & -i\sigma^j\\
		i\sigma^j & 0\\
	\end{pmatrix}\hspace{1cm}(j=1,2,3).
\end{align}
 $\Gamma_\alpha$ represents the spin connection coefficients, which are defined as
\begin{align}\label{eqn 6}
	\Gamma_\alpha=\dfrac{1}{8}[\gamma^a,\gamma^b]e^\nu_a e_{b\nu;\alpha} ,
\end{align}
where $e_{b\nu;\alpha}=\partial_{\alpha} e_{b\nu}-\Gamma^\beta_{\alpha\nu}$.
 We choose the tetrad from Eq. \eqref{eqn 1} as
 \begin{align}\label{eqn 7}
 	e^a_\alpha= diag \left( \sqrt{A},\dfrac{\sqrt{1+L}}{\sqrt{A}},r,r\sin\theta \right).
 \end{align}
 The spin connection coefficients from Eq. \eqref{eqn 6} can be calculated as
 \begin{align}\label{eqn 8}
 &\Gamma_0=-\dfrac{1}{4}\dfrac{1}{\sqrt{1+L}}\dfrac{dA}{dr}\gamma_0\gamma_1,\hspace{1cm}\Gamma_1=0,
 \nonumber\\ 
&\Gamma_2=\dfrac{1}{2}\dfrac{\sqrt{A}}{\sqrt{1+L}}\gamma_1\gamma_2,\hspace{1cm} \Gamma_3=\dfrac{1}{2}\left(\sin \theta \dfrac{\sqrt{A}}{\sqrt{1+L}}\gamma_1\gamma_3+\cos\theta\gamma_2\gamma_3 \right).
\end{align}
Using Eqs. \eqref{eqn 6}-\eqref{eqn 8} in Eq. \eqref{eqn 4}, we get
 \begin{align}\label{eqn 9}
 	&\dfrac{\gamma_0}{\sqrt{A}}\dfrac{d \Psi}{d t}-\dfrac{\gamma_1 \sqrt {A}}{\sqrt{1+L}}\left(\dfrac{d}{d r}+\dfrac{1}{r}+\dfrac{1}{4A}\dfrac{d A}{d r}\right)\Psi-\dfrac{\gamma_2}{r}\left(\dfrac{d}{d \theta}+\dfrac{1}{2}\cot\theta\right)\Psi -\dfrac{\gamma_3}{r\sin\theta}\dfrac{d\Psi}{d\phi}-m\Psi=0.
 \end{align}
  The wave function $\Psi$ in the above equation can be taken as
  \begin{align} \label{eqn 10}
  \Psi= A^{-{1}/{4}}\Phi,
\end{align}   
the wave Eq. (\ref{eqn 9}) reduces to               
\begin{align}\label{eqn 11}
 	\dfrac{\gamma_0}{\sqrt{A}}\dfrac{d\Phi}{dt}-\dfrac{\gamma_1 \sqrt {A}}{\sqrt{1+L}}\left(\dfrac{d }{d r}+\dfrac{1}{r}\right)\Phi-\dfrac{\gamma_2}{r}\left(\dfrac{d}{d \theta}+\dfrac{1}{2}\cot\theta\right)\Phi-\dfrac{\gamma_3}{r\sin\theta}\dfrac{d \Phi}{d \phi}-m\Phi=0.
 \end{align}
  It is noted that there exist two different spin magnetic quantum numbers. We need to define the wave function separately. Therefore the ansatz can be taken as
 \begin{align}\label{eqn 12}
 	\Phi(t,r,\theta,\phi)= \begin{pmatrix}
   &\dfrac{ig^\pm(r)}{r}\phi^\pm_{jm}(\theta,\phi)\\
   &\dfrac{f^\pm(r)}{r}\phi^\mp_{jm}(\theta,\phi)
    \end{pmatrix}e^{-i\omega t},
 	\end{align}\\
where
 	\begin{align}
       & \phi^+ _{jm}= \begin{pmatrix}
       &\sqrt{\dfrac{{j+m}}{2j}}Y^{m-1/2}_l \\
       &\sqrt{\dfrac{{j-m}}{2j}}Y^{m+1/2}_l
               \end{pmatrix},\hspace{0.4cm}     \left({\rm  for }~ j= l+\dfrac{1}{2}\right)
               \nonumber\\	
 		&\phi^-_{jm}= \begin{pmatrix}
 			&\sqrt{\dfrac{{j+1-m}}{2j+2}}Y^{m-1/2}_l \\
 			&-\sqrt{\dfrac{{j+1+m}}{2j+2}}Y^{m+1/2}_l
 		\end{pmatrix}, \hspace{0.4cm} \left({\rm for}~j= l-\dfrac{1}{2} \right).
   	\end{align}
Using the tortoise coordinate \eqref{eqn 14} and substituting Eq. (\ref{eqn 12}) in Eq. (\ref{eqn 11}), we can write the Dirac equation into simplified matrix form as
 	
 	\begin{align}\label{eqn 15}
    -\begin{pmatrix}
 			& 0 & -\omega \\
 			& \omega & 0 
 		\end{pmatrix}\begin{pmatrix}
 		& f^\pm\\
 		& g^\pm
 		\end{pmatrix}+ \dfrac{d}{d r_*}
 	\begin{pmatrix}
 	& f^\pm\\
 	& g^\pm
 	\end{pmatrix} -\sqrt{A}\begin{pmatrix}
 	&\dfrac{\kappa_\pm}{r} & m \\
 	& m & -\dfrac{\kappa_\pm}{r}
 \end{pmatrix}\begin{pmatrix}
 	& f^\pm\\
 	& g^\pm
 \end{pmatrix} =0,
 \end{align}
 where the constants $\kappa_{\pm}$ denote the negative and positive integers, which are given by 
\begin{align}
    \kappa\pm = \begin{pmatrix}
    & j+\dfrac{1}{2}, & j= l+\dfrac{1}{2} \\
    \nonumber\\
    &-\left ( j+{\dfrac{1}{2}}\right), & j= l-\dfrac{1}{2}
    \end{pmatrix}.
\end{align}
Since the Eq. (\ref{eqn 1}) is a spherically symmetric black hole, we shall consider the radial functions ($g^{\pm}$ and $f^{\pm}$). According to Chandrasekhar \cite{chandrasekhar1983} the variables can be transformed to 
\begin{align}\label{eqn 16}
\begin{pmatrix}
 \hat f^\pm \\
  \hat g^\pm
\end{pmatrix}=\begin{pmatrix}
& \sin\left(\dfrac{\theta_{\pm}}{2}\right)& \cos\left(\dfrac{\theta_{\pm}}{2}\right)\\
&  \cos\left(\dfrac{\theta_{\pm}}{2}\right)& -\sin\left(\dfrac{\theta_{\pm}}{2}\right)
\end{pmatrix}\begin{pmatrix}
& f^\pm \\
& g^\pm
\end{pmatrix},
\end{align}
where\hspace{.5cm}$\theta_+ = tan^{-1} \left({mr}/{|\kappa_{\pm}|}\right)$. After some calculations, Eq. (\ref{eqn 15}) reduces to  \\
\begin{align}\label{eqn 17}
&\dfrac{d}{d{r_*}}\begin{pmatrix}
 \hat f^\pm \\
 \hat g^\pm
\end{pmatrix}-\sqrt{A}\sqrt{\left(\dfrac{\kappa_\pm}{r}
\right)^2+m^2}\begin{pmatrix}
& 1 & 0 \\
& 0 & -1 
\end{pmatrix}
\begin{pmatrix}
 \hat f^\pm \\
 \hat g^\pm
\end{pmatrix}= -\omega\left(1+\dfrac{A}{2\omega\sqrt{1+L}}\dfrac{m |\kappa_\pm|}{\kappa^2_{\pm}+m^2r^2}\right)
\begin{pmatrix}
& 0 & -1 \\
& 1 & 0 
\end{pmatrix}
\begin{pmatrix}
 \hat f^\pm \\
 \hat g^\pm
 \end{pmatrix}.
\end{align} 
If we choose the transformation $\hat r _* = r_* + \rm tan ^{-1}\left({mr}/{|\kappa_\pm|}\right)/{2\omega}$, the above  equation reduces to
 \begin{align} \label{eqn 18}
 -\dfrac{d}{d{\hat r_*}}\begin{pmatrix}
 \hat f^\pm \\
 \hat g^\pm
\end{pmatrix}+ W_{\pm}\begin{pmatrix}
 -\hat f^\pm \\
 \hat g^\pm
\end{pmatrix}= \omega\begin{pmatrix}
 \hat g^\pm \\
 \hat f^\pm
\end{pmatrix},
\end{align}

where
\begin{align}\label{eqn 19}
&W_\pm 
=\dfrac{\sqrt{A}\sqrt{\kappa^2_\pm + m^2r^2}}{r  P_{\pm}}, 
~~~P_{\pm}= \left(1\pm \dfrac{A}{2\omega\sqrt{1+L}}\dfrac{m \vert \kappa_\pm \vert}{\kappa^2_{\pm}+m^2r^2 }\right).
\end{align}
We can write the radial decouple Dirac equations as follows
\begin{align}\label{eqn Vdm1}
\left(-\dfrac{d^2}{d\hat{r}^2_*}+V_{\rm dm\pm}\right)\hat f^\pm =\omega^2\hat f^\pm ,
~~~
\left(-\dfrac{d^2}{d\hat{r}^2_*}+V_{\rm dm\pm}\right)\hat g^\pm =\omega^2\hat g^\pm ,
\end{align}
where $V_{dm\pm}$ is the effective potential of massive Dirac field perturbation which is given by 
\begin{align}\label{eqn Vdm}
 V_{ dm \pm} = & W^2_\pm \pm \dfrac{d{W_\pm}}{d\hat r_*} .\nonumber\\
       =&\left(\dfrac{\sqrt{A}\sqrt{\kappa^2 _{\pm}  + m^2r^2}}{r  P_{\pm}}\right)^2 \pm \dfrac{A}{\sqrt{1+L}P_\pm}\left[ - \dfrac{\kappa^2_\pm \sqrt{A}}{P_\pm r^2 \sqrt{\kappa_{\pm}^2+m^2r^2}} +  \dfrac{\sqrt{\kappa ^2 _\pm+ m^2 r^2} A'}{2 r \sqrt{A}P_\pm} 
 - \dfrac{\sqrt{\kappa_{\pm}^2 + m^2r^2} \sqrt{A}}{r P_{\pm}^2}   \right. \nonumber\\ & \left. \times \left( - \dfrac{\kappa_\pm m^3 r A}{\omega \sqrt{1+L} (\kappa^2_{\pm}+m^2r^2)^2} +  \dfrac{\kappa_{\pm} m  A'}{2\omega \sqrt{1+L} (k^2_{\pm}+m^2r^2)}\right)  
\right] .
\end{align}
The effective potential of massless Dirac perturbation is obtained as
\begin{align}
V_{d\pm}=\dfrac{A \kappa^2_{\pm}}{r^2} \pm \left[	\dfrac{\kappa_\pm \sqrt{A} A'}{2 r \sqrt{1+L}}-\dfrac{\kappa_\pm A^\frac{3}{2}}{r^2 \sqrt{1+L}}	\right].
\end{align}

\begin{figure}[h!]
\centering
  \subfloat[\centering ]{{\includegraphics[width=170pt,height=170pt]{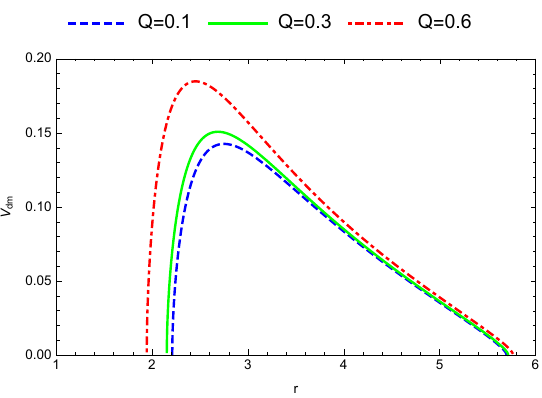}}\label{fig 3a}}
  \qquad
   \subfloat[\centering ]{{\includegraphics[width=170pt,height=170pt]{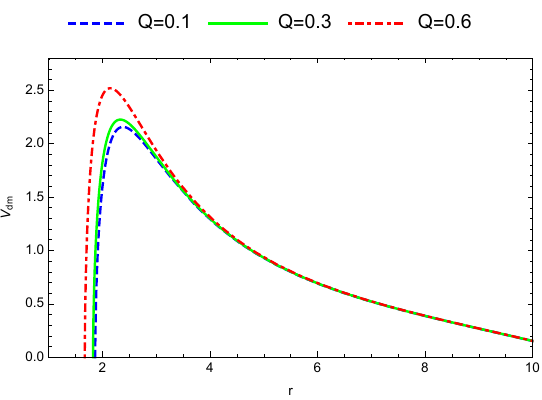}}\label{fig 3b}}
   \caption{Variation of the effective potential for the massive Dirac field for different values of $Q$. The physical parameters are chosen as (a) $M=1$, $\Lambda=0.05$, $L=0.2$, $\ell=1$ \text{and} $m=0.1$ (b) $M=1$, $\Lambda=-0.05$, $L=0.2$, $\ell=1$ \text{and} $m=0.1$.}
   \label{fig 3}
\end{figure}

\begin{figure}[h!]
\centering
  \subfloat[\centering ]{{\includegraphics[width=170pt,height=170pt]{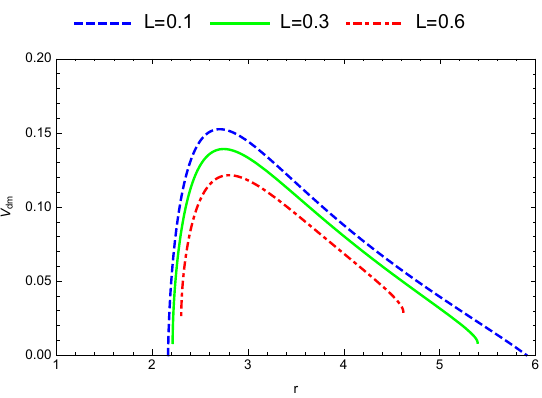}}\label{fig 4a}}
  \qquad
   \subfloat[\centering ]{{\includegraphics[width=170pt,height=170pt]{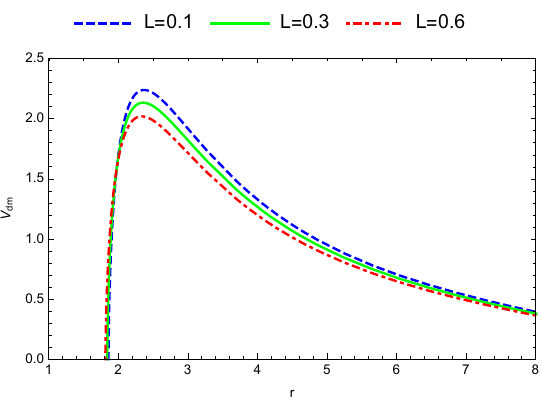}}\label{fig 4b}}
   \caption{Variation of the effective potential for the massive Dirac field for different values of $L$. The physical parameters are chosen as (a) $M=1$, $\Lambda=0.05$, $Q=0.2$, $\ell=1$ \text{and}  $m=0.1$ (b) $M=1$, $\Lambda=-0.05$, $Q=0.2$, $\ell=1$ \text{and}  $m=0.1$.}
   \label{fig 4}
\end{figure}

 It is noted that the Dirac particles and  
antiparticles have same QNMs for spherically symmetric black hole spacetime. Therefore the function $\hat{f}^{+}$ represents all the connected physics of Dirac field perturbation in such black holes. Figs. \ref{fig 3}  and  \ref{fig 4} indicate the graphs of effective potentials of Dirac field perturbation for different values of $Q$ and $L$ with the fix of other black hole parameters.
Increasing charge $Q$ increases the peak of effective potentials for both RNdS and RNAdS-like black holes. However, the peak of effective potential decreases with increasing the parameter $L$ for both RNdS and RNAdS-like black holes.   Thus the nature of influences of $Q$ and $L$ on the peak of the effective potential for RNdS  black hole remains consistent across all types of perturbations. However, for RNAdS the influence of $L$  on the effective potential varies with the types of perturbations: with increasing $L$ the peak of effective potentials for scalar and electromagnetic field perturbations increase but for Dirac field perturbation it decreases. In the following section, we shall consider the greybody factor,  QNMs for RNdS and RNAdS-like black holes in bumblebee gravity.

\section{Greybody factor} \label{sec 5}
Hawking showed \cite{hawking1975} that the black hole emits radiation due to vaccum fluctuation near the even horizon of black hole. The radiation emitted from the black hole is known as Hawking radiation. When Hawking radiations evaporate out from the black hole event horizon, it encounters the spacetime curvature generated by its black hole source. Therefore an observer located at infinite distance will observe the modified form of thermal radiation which is distinct from the original thermal radiation near the event horizon. The difference of modified and original thermal radiations may be calculated by the so-called greybody factor.
 Applying  the general semi-analytic bounds of greybody factor given in \cite{M. Visser1999,P. Boonserm2008}, we will investigate the greybody factors of Dirac, Scalar and electromagnetic fields emitted from the RNdS-like black holes.
The rigorous bound on greybody factor is given by
\begin{align}\label{eqn gb1}
T \geq \sech^2\left( \dfrac{1}{2\omega}\int _{-\infty}^{\infty}\wp d{r_*}\right),
\end{align}
where 
\begin{align}
\wp=\dfrac{1}{2h(r_*)}\sqrt{[h'(r_*)]^2+(\omega^2-V-h^2(r_*))^2}.
\end{align}
Here $h(r_*)$ is a positive function and it must hold the conditions $h(+\infty)$=$h(-\infty)$=$\omega$ \cite{M. Visser1999}. One can simply set $h=\omega$. The presence of the cosmological horizon and the event horizon of RNdS-like  black hole give a distinct set-up for greybody factor which is based on wave scattering between two horizons. But RNAdS-like black hole gives time like boundary at spatial infinity that needs a careful treatment of boundary conditions. Applying rigorous bound technique, the study of greybody factor for RNAdS-like black hole is very difficult.  
In this paper the  study of greybody factor restricted for RNdS black hole. Thus Eq. \eqref{eqn gb1} reduces to 
\begin{align}\label{eqn gb2}
T \geq \sech^2\left( \int _{-\infty}^{\infty}\dfrac{V}{2\omega}d{r_*}\right).
\end{align}
For dS case, using the definition of $r_*$, the relation for bounds on greybody factors can be written as 
\begin{align}\label{eqn gb3}
T \geq \sech^2\left( \dfrac{\sqrt{1+L}}{2\omega}\int _{r_h}^{r_c}\dfrac{V}{A(r)} d{r}\right).
\end{align}
\subsection{Scalar and electromagnetic perturbations}

\begin{figure}[h!]
\centering
  \subfloat[\centering ]{{\includegraphics[width=170pt,height=170pt]{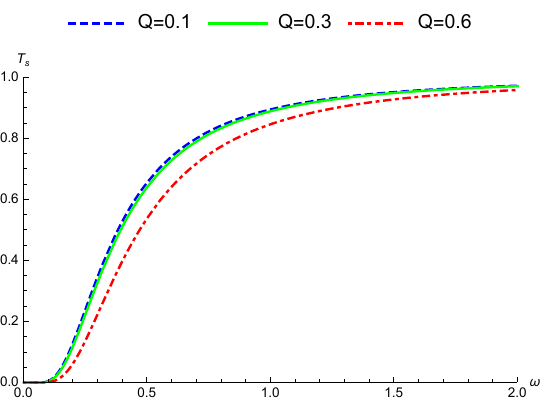}}\label{fig 9a}}
  \qquad
   \subfloat[\centering ]{{\includegraphics[width=170pt,height=170pt]{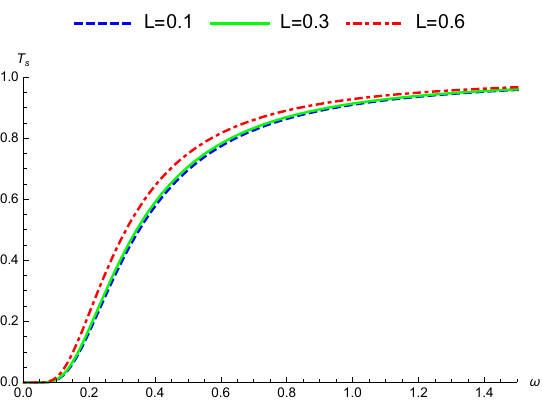}}\label{fig 9b}}
   \caption{Variation of the Greybody factor for the massive scalar field (a) for different values of $Q$ with fixed $M=1,\Lambda=0.05,$ $m=0.1$, $s=0$, $\ell=1$ \text{and} $L=0.2$ ; (b) for different values of $L$ with fixed $M=1, \Lambda=0.05, $ $m=0.1$, $s=0$, $\ell=1$ \text{and} $Q=0.2$.}
   \label{fig 9}
\end{figure}

\begin{figure}[h!]
\centering
  \subfloat[\centering ]{{\includegraphics[width=170pt,height=170pt]{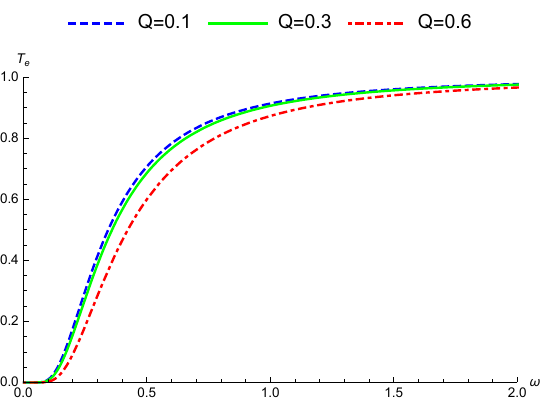}\label{fig 10A}}}
  \qquad
   \subfloat[\centering ]{{\includegraphics[width=170pt,height=170pt]{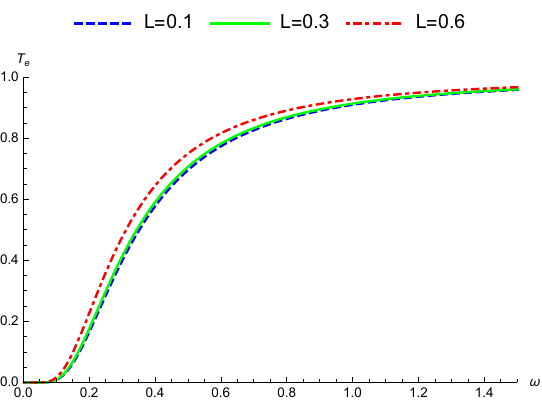}}\label{fig 10B}}
   \caption{Variation of the Greybody factor for the electromagnetic field (a) for different values of $Q$ with fixed $M=1, \Lambda=0.05$,  $s=1$, $\ell=1$ \text{and} $L=0.2$ ; (b) for different values of $L$ with fixed $M=1$,  $\Lambda=0.05$,  $s=1$, $\ell=1$ \text{and} $Q=0.2$.}
   \label{fig 10.}
\end{figure}

Applying  Eqs. (\ref{eqn Vs}) and (\ref{eqn Ve}) in Eq. (\ref{eqn gb3}), the expression of the greybody factor for scalar perturbation and electromagnetic  perturbation of RNdS black hole can be recast as
\begin{align}
T_{s,e}\geq  \sech^2 \left(\frac{\sqrt{1+L}}{2\omega}\int_{r_h}^{r_c} A\left[\dfrac{l(1+l)}{r^2}+(1-s^2)\left(  \dfrac{1}{r(1+L)}\dfrac{d{A}}{d{r}}+m^2\right) \right]dr\right).
\end{align}
The rigorous bounds of RNdS-like black hole for the scalar and electromagnetic perturbations are derived as 

\begin{align}\label{eqn greyse}
T_{s,e} \geq  & \sech^2\left[ \frac{\sqrt{1+L}}{2\omega}\left\lbrace (1-s^2)(m^2-\frac{2}{3}\Lambda) (r_c-r_h)-l(1+l)\left(\frac{1}{r_c}-\frac{1}{r_h}\right)- 
\frac{M(1-s^2)}{(1+L)}\left(\frac{1}{r_c^2}-\frac{1}{r_h^2}\right)  \right.\right. \nonumber\\ & \left.\left.+\frac{4(1-s^2)Q^2}{3Q^2(2+L)}\left(\frac{1}{r_c^3}-\frac{1}{r_h^3}\right)        \right\rbrace 
\right].
\end{align}
It is noted from Eq. \eqref{eqn greyse} that both the lower bounds of the greybody factor of scalar field and electromagnetic field perturbations depend not only on the Lorentz violation parameter but also on the difference between the two horizons.
The graphs of rigorous bound of the greybody factor for scalar field perturbation and electromagnetic field perturbation are drawn in Figs. \ref{fig 9a}, \ref{fig 9b} and Figs. \ref{fig 10A}, \ref{fig 10B} respectively for different values of $Q$ and $L$.
Increasing $Q$ decreases the bound of the greybody factor for both scalar and electromagnetic filed perturbations of RNdS-like black hole, which leads to lesser wave to reach a far distant observer. However, it has an opposite effect in the greybody factor for both scalar and electromagnetic field perturbations of RNdS-like black hole with  the increase in $L$. Therefore increasing $L$ gives higher probability for radiation to escape to distant observers.

\subsection{Greybody factor for Dirac particle}
	In this  section, we discuss the greybody factor for massless and massive fermions of RNdS-like black hole. Using Eq. (\ref{eqn Vdm}),  Eq. (\ref{eqn gb3}) reduces to 
\begin{align}\label{eqn gbVdm}
T \geq \sech^2\left( \dfrac{1}{2\omega}\int _{r_h}^{r_c}\left | \dfrac{dW}{dr_*} \right | dr_*+\dfrac{1}{2\omega}\int _{r_h}^{r_c}\left|W^2\right|dr_* \right). 
\end{align}
We shall consider the first and second integral in Eq. (\ref{eqn gbVdm}) separately. The first integral may be calculated as follows 
\begin{align}\label{eqn 36}
\int _{r_h}^{r_c}\left | \dfrac{dW}{dr_*} \right | dr_*= W|^{r_c}_{r_h}=0.
\end{align}
It is noted that $W$ approaches  zero at the horizons. The contribution of rigorous bound may be derived from the  second integral only. Completing the integral given in Eq. (\ref{eqn gbVdm}), we obtain
\begin{align}\label{eqn 37}
\int_{r_h}^{r_c}|W^2|dr_*=\int_{r_h}^{r_c}\frac{1}{r^2} \dfrac{2\omega(1+L)(\kappa^2+m^2r^2)^2}{2\omega\sqrt{1+L}(\kappa^2+m^2r^2)+A m \kappa}dr.
\end{align}
The findings of greybody factors for the massless and the massive fermions will be significantly different. Hence we discuss the calculation one by one.

\subsubsection{Massless fermion}\label{eqn gb ml}
\begin{figure}[h!]
\centering
  \subfloat[\centering ]{{\includegraphics[width=170pt,height=170pt]{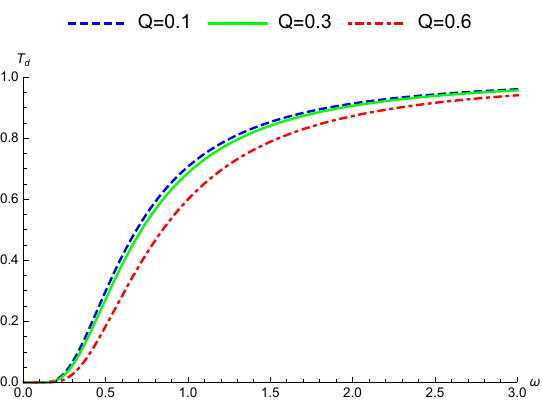}}\label{fig 11a }}
  \qquad
   \subfloat[\centering ]{{\includegraphics[width=170pt,height=170pt]{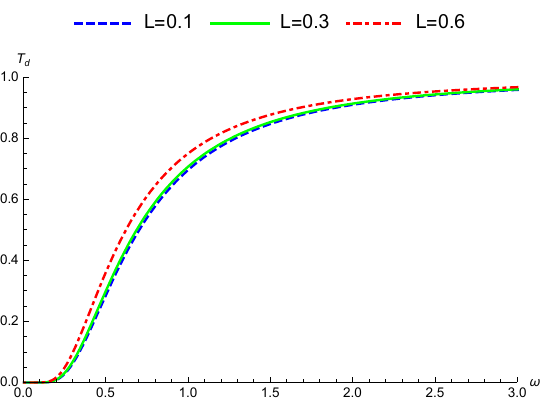}}\label{fig 11b}}
   \caption{Variation of the Greybody factor for the massless Dirac field (a) for different values of $Q$ with fixed $M=1$,      $\Lambda=0.05$, $\ell=0$ \text{and} $L=0.2$ ; (b) for different values of $L$ with fixed $M=1$,  $\Lambda=0.05$, $\ell=0$ \text{and} $Q=0.2$.}
   \label{fig 11}
\end{figure}

If we take $m=0$ in Eq. (\ref{eqn 37}), the integral can be calculated as
\begin{align}
\int_{r_h}^{r_c}|W^2|dr_*=\int_{r_h}^{r_c}\dfrac{\kappa^2\sqrt{1+L}}{r^2}   dr =  \kappa^2\sqrt{1+L} \left( \frac{1}{r_h}-\frac{1}{r_c}\right).
\end{align}
Using Eq. (\ref{eqn gbVdm}), we derive the rigorous bound as  
\begin{align}\label{eqn gbmassless}
T_{d}= \sech^2\left[\frac{\kappa^2\sqrt{1+L}}{2\omega}\left(\frac{r_c-r_h}{r_h r_c}\right)\right].
\end{align}
We see that the bound depends not only on $L$ but also on the distance between the horizons. This shows that the greybody factor bound is dependent on $L$ and model parameter of the different horizons.   From Fig. \ref{fig 11}, we observe that the greybody factor of massless Dirac particle increases with increasing $L$ and decreases with increasing $Q$. Thus increasing the value of $L$ allows more waves to be transmitted to the far observer. Further, the greybody factor  decreases with increasing the distance between the horizons. One can clearly see from Fig. \ref{fig 1} and \ref{fig 2} that the distance between the horizons increases with increasing $Q$ or decreasing $L$, this behavior is consistent with  with the trends of greybody factor observed in Figs. \ref{fig 11}.


\subsubsection{Massive Dirac field} 
\begin{figure}[h!]
\centering
  \subfloat[\centering ]{{\includegraphics[width=170pt,height=170pt]{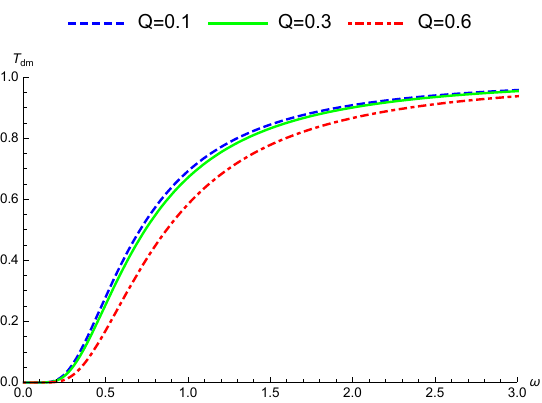}}\label{fig 12a}}
  \qquad
   \subfloat[\centering ]{{\includegraphics[width=170pt,height=170pt]{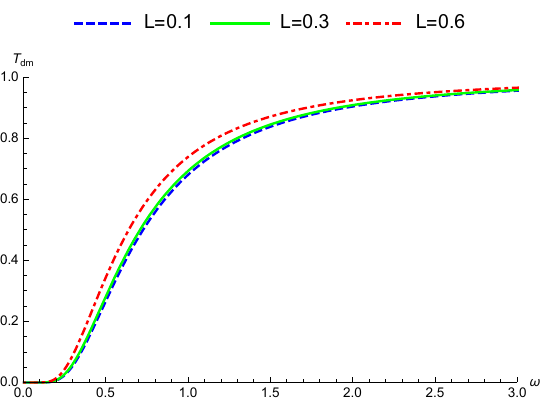}}\label{fig 12b}}
   \caption{Variation of the Greybody factor for the massive Dirac field (a) for different values of $Q$ with fixed $M=1$,      $\Lambda=0.05$, $m=0.1$, $\ell=1$ and  $L=0.2$ ; (b) for different values of $L$ with fixed $M=1$,  $\Lambda=0.05$, $m=0.1$, $\ell=1$ \text{and} $Q=0.2$.}
   \label{fig 12}
\end{figure}

For the massive fermion, we have to find the total integration in Eq. (\ref{eqn 37}). The result can be analytically calculated but it is very tedious to investigate the behaviour of the greybody factor. The calculation and the results  have been displayed in the Appendix A. To study greybody factor bound, the approximation method will be used in this section. First, Eq. \eqref{eqn 37} is reduces to

\begin{align}\label{eqn gbm 1}
T \geq & \sech^2 \dfrac{1}{2\omega}\int_{r_h}^{r_c}\frac{1}{r^2}\dfrac{\kappa^2\sqrt{1+L}(1+ \tilde{\mu}^2 r^2)}{(1+\frac{A \tilde{\mu }}{2\omega\sqrt{1+L}(1+\tilde{\mu}^2r^2)}) }dr \nonumber\\ & =  \sech^2 \dfrac{1}{2\omega}\int^{r_c}_{r_h} \beta dr = T_{dm},
\end{align}
where
\begin{align}
\beta =\frac{1}{r^2} \dfrac{\kappa^2\sqrt{1+L}(1+ \tilde{\mu}^2 r^2)}{\left(1+\frac{A \tilde{\mu} }{2\omega\sqrt{1+L}(1+\tilde{\mu}^2r^2)}\right)}, \hspace{.5cm}
\tilde{\mu}=\frac{m}{\kappa}.
\end{align}
It is noted that the factor $\left(1+\frac{A \tilde{\mu} }{2\omega\sqrt{1+L}(1+\tilde{\mu}^2r^2)}\right)$ is bigger than 1 in the integrand $\beta$. We can use this inequality to approximate a new integrand, which leads to the new greybody factor bound. Hence, the integrand $\beta$ can be expressed as

\begin{align}\label{eqn gbm 2}
\beta= \frac{1}{r^2} \dfrac{\kappa^2\sqrt{1+L}(1+ \tilde{\mu}^2 r^2)}{\left(1+\frac{A \tilde{\mu} }{2\omega\sqrt{1+L}(1+\tilde{\mu}^2r^2)}\right)}\leq \dfrac{\kappa^2}{r^2}\sqrt{1+L}(1+\tilde{\mu}^2 r^2)= \beta_{app}.
\end{align}
We know that $\beta$ and $\beta_{app}$ are positive functions in the range $r_h < r < r_c $. Then the integral may be written as $\int \beta dr$ $\leq$  $\int \beta _{app}~dr$. We derive the greybody bound as follows 
\begin{align}\label{eqn gbm 3}
T\geq \sech^2 \left(\frac{1}{2\omega}\int_{r_h}^{r_c} \beta dr \right)\geq \sech^2 \left(\frac{1}{2\omega}\int_{r_h}^{r_c} \beta_{app} dr \right)= T_{dm}.
\end{align}
Completing the integral in Eq. \eqref{eqn gbm 3}, we get
\begin{align}\label{eqn gbm}
 T_{dm}&=\sech^2 \left(\frac{1}{2\omega}\int_{r_h}^{r_c} \beta_{app} dr \right) \nonumber \\
&=\sech^2 \left[ \frac{\sqrt{1+L}}{2\omega} \dfrac{(r_c - r_h)\kappa^2}{r_c  r_h}\left(1+\frac{m^2}{\kappa^4}  r_c r_h
\right)\right].
\end{align}
The rigorous bounds on the greybody factor of massive fermion drawn in Figs. \ref{fig 12a} and \ref{fig 12b} for varying $Q$ and $L$. The variation of the rigorous bounds on the greybody factor for massive fermion indicates a similar behaviour to that of scalar and electromagnetic field perturbations. The greybody factor decreases with the increase  of $Q$ but  increases with the increase of $L$.

\section{Greybody factor of massless perturbations in different black holes}\label{sec 6}
\begin{figure}[h!]
\centering
  \subfloat[\centering ]{{\includegraphics[width=170pt,height=170pt]{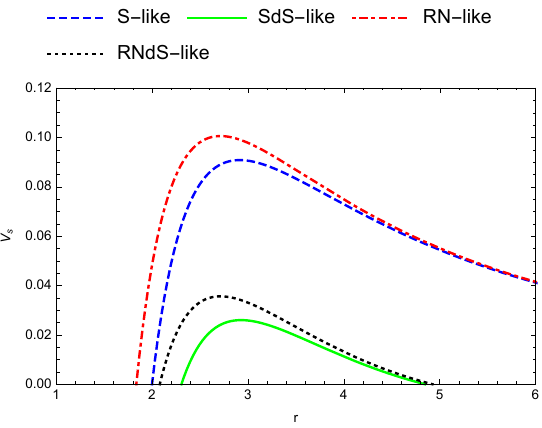}}\label{fig 13a}}
  \qquad
   \subfloat[\centering ]{{\includegraphics[width=170pt,height=170pt]{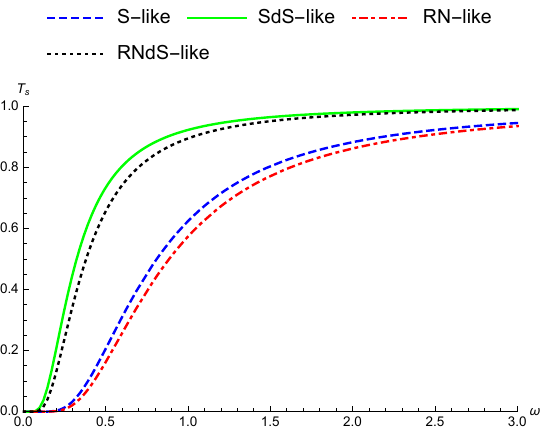}}\label{fig 13b}}
   \caption{ Variation of  (a) effective potential for the scalar field,  with fixed  $M=1$, $\Lambda=0.05$, $\ell=1$ \text{and} $L=0.2$ (b) greybody factor for the scalar field with fixed $M=1$, $\Lambda=-0.05$, $\ell=1$ \text{and} $L=0.2$.}
   \label{fig 13}
\end{figure}

\begin{figure}[h!]
\centering
  \subfloat[\centering ]{{\includegraphics[width=170pt,height=170pt]{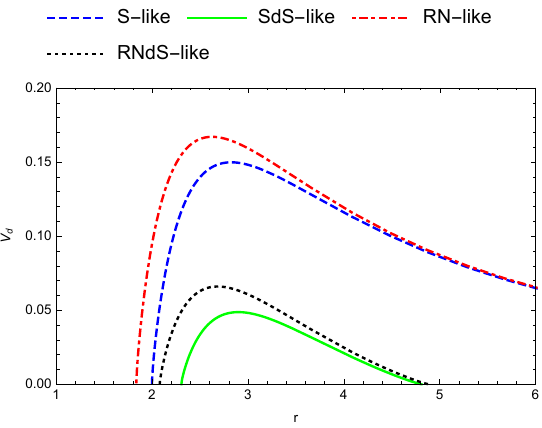}}\label{fig 14a}}
  \qquad
   \subfloat[\centering ]{{\includegraphics[width=170pt,height=170pt]{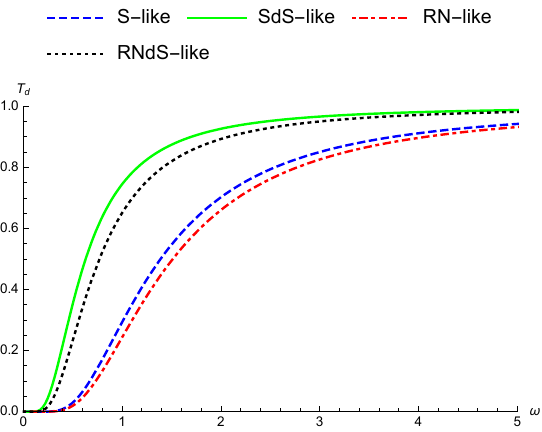}}\label{fig 14b}}
   \caption{ Variation of (a) effective potential for the Dirac  field, with fixed $M=1$, $\Lambda=0.05$, $\ell=1$ \text{and} $L=0.2$, (b) greybody factor for the Dirac field, with fixed $M=1$, $\Lambda=-0.05$, $\ell=1$ \text{and} $L=0.2$.}
   \label{fig 14}
\end{figure}

\begin{figure}[h!]
\centering
  \subfloat[\centering ]{{\includegraphics[width=170pt,height=170pt]{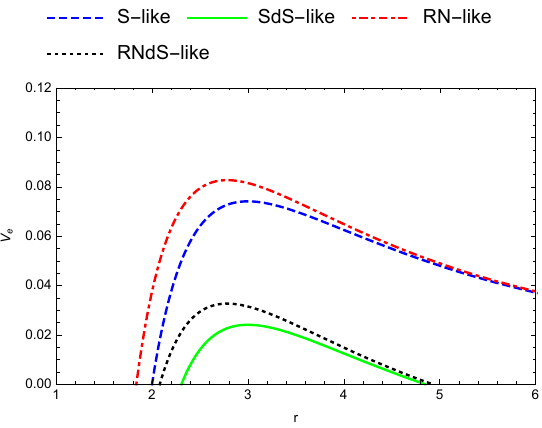}}\label{fig 15a}}
  \qquad
   \subfloat[\centering ]{{\includegraphics[width=170pt,height=170pt]{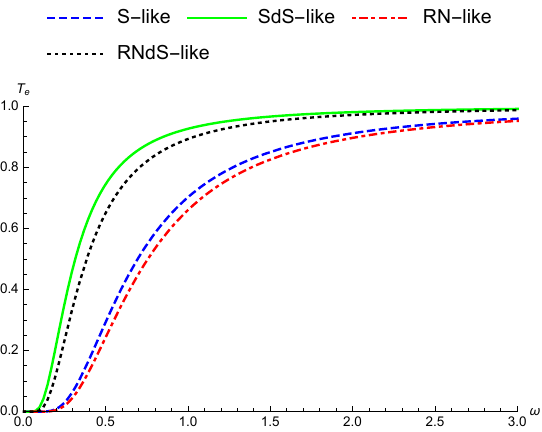}}\label{fig 15b}}
   \caption{ Variation of  (a) effective potential for the electromagnetic field, with fixed $M=1$, $\Lambda=0.05$, $\ell=1$ \text{and} $L=0.2$, (b) greybody factor for the electromagnetic field with fixed $M=1$, $\Lambda=-0.05$, $\ell=1$ \text{and} $L=0.2$.}
   \label{fig 15}
\end{figure}

In this section, we can compare the greybody factor of massless perturbations for  different kinds of black holes namely Schwarzchild-like (S),  Schwarzchild-de Sitter-like (SdS), Reissner-Norstorm-like (RN) and RNdS-like black holes. It is noted that $r_c$ tends to infinity  in Eq. \eqref{eqn greyse} and \eqref{eqn gbmassless} for the S-like black hole and RN-like black hole \cite{ngampitipan2013}. In such case the argument of function $\sech$ of RN-like black hole is bigger than other black holes in all the perturbations. Hence the RN-like black hole has minimum greybody factor  than other black holes in all the perturbations as shown in the right panels of the Figs. \ref{fig 13}-\ref{fig 15}. This implies that the effective potential for RN-like black hole is larger than other black holes as given in the left panels of Figs. \ref{fig 13}-\ref{fig 15}. For the RNdS-like and SdS-like black holes, the horizons between $r_h$ and $r_c$ are closer and thinner due to presence of cosmological constant $\Lambda$. The presence of $Q$ in RNdS-like black hole, the argument of function $\sech$ is slightly bigger than SdS-like black hole. As a result, more waves may transmit through the potential for the SdS-like black hole than RNdS-like black hole so that the greybody factor of  SdS-like black hole is bigger than the greybody factor of RNdS-like black hole as shown in the right panels in Figs. \ref{fig 13}-\ref{fig 15}. For the functioning of greybody bound in SdS-like black hole and RNdS-like black hole, we may take $\delta{r}= r_c- r_h $ because the argument of function $\sech$ relies on the distance between the two horizons. If $\delta{r}$ is very small, the greybody factor bound will be large and when $\delta r$ is very large, the greybody factor bound  will be small for all perturbations.


\section{Quasinormal modes (QNMs)}\label{sec 7}
\begin{figure}[h!]
\centering
  \subfloat[\centering ]{{\includegraphics[width=160pt,height=160pt]{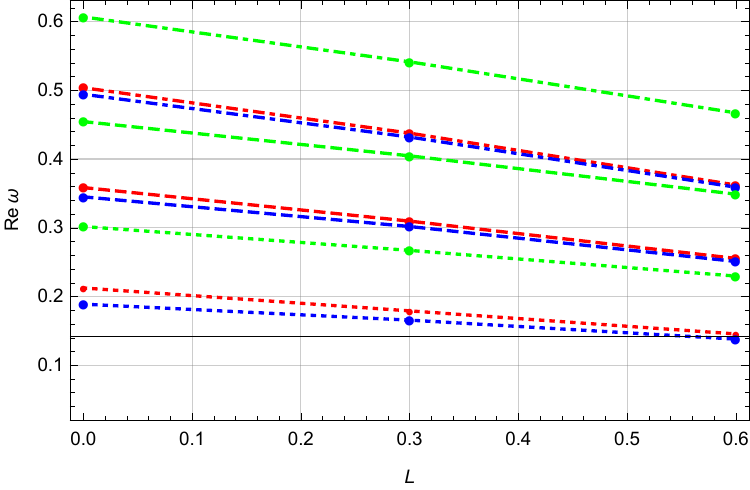}}\label{fig realL}}
  \quad \quad
   \subfloat[\centering ]{{\includegraphics[width=160pt,height=160pt]{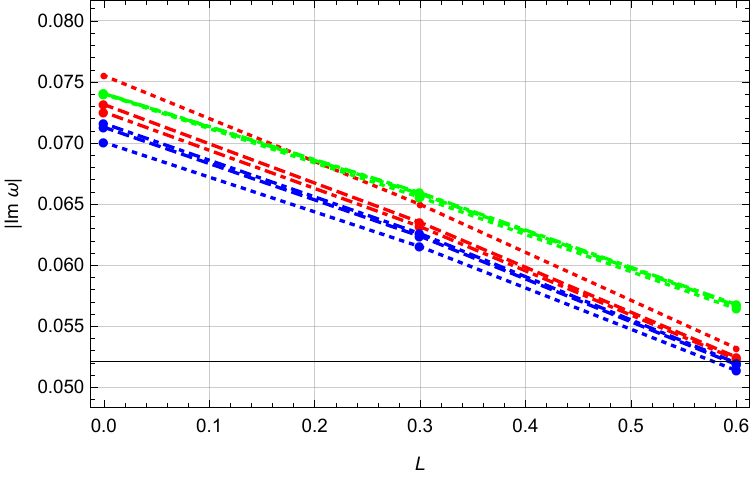}}\label{fig imaginaryL}}
   \caption{Variation of real and imaginary parts of the quasinormal frequencies for scalar, electromagnetic and Dirac perturbations for different values of $L$ with fixed $M=1$, $\Lambda=0.05$ and $Q=0.2$. The colours corresponding to scalar, electromagnetic and Dirac perturbations are red, blue and green respectively. The dotted, dashed and dot-dashed represent the frequencies of $(l=1,n=0)$, $(l=2,n=0)$ and $(l=3,n=0)$ respectively. }
   \label{fig qnmL}
\end{figure}

\begin{figure}[h!]
\centering
  \subfloat[\centering ]{{\includegraphics[width=160pt,height=160pt]{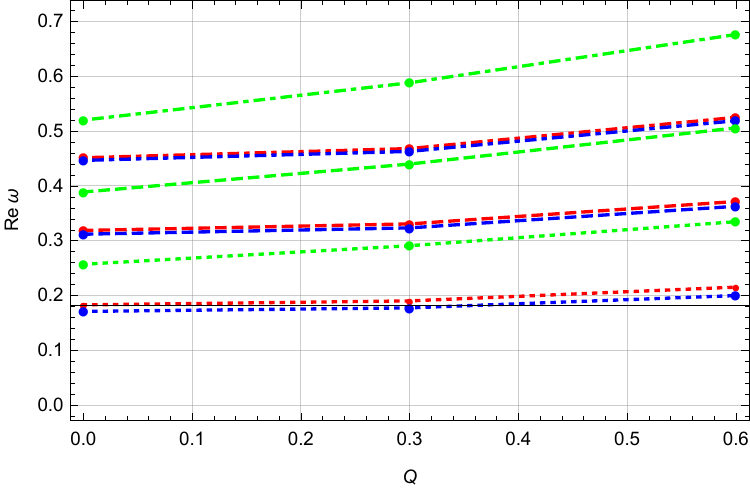}}\label{fig realQ}}
  \quad \quad
   \subfloat[\centering ]{{\includegraphics[width=160pt,height=160pt]{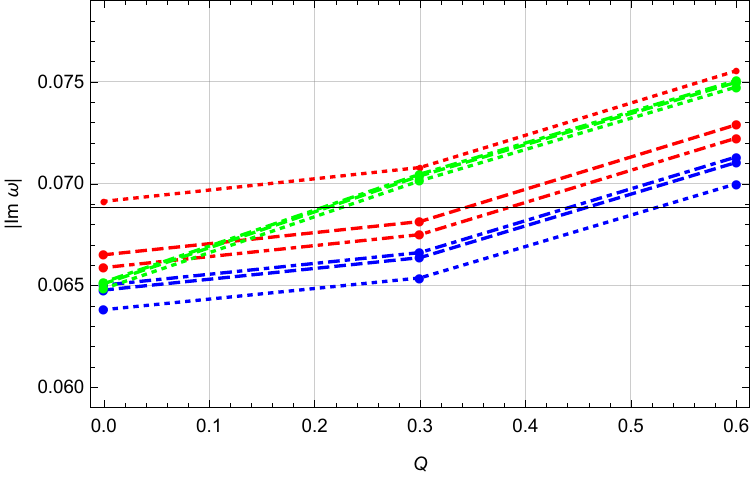}}\label{fig imaginaryQ}}
   \caption{Variation of real and imaginary parts of the quasinormal frequencies for scalar, electromagnetic and Dirac perturbations for different values of $Q$ with fixed $M=1$, $\Lambda=0.05$ and $L=0.2$. The colours corresponding to scalar, electromagnetic and Dirac perturbations are red, blue and green respectively. The dotted, dashed and dot-dashed represent the frequencies of $(l=1,n=0)$, $(l=2,n=0)$ and $(l=3,n=0)$ respectively.}
   \label{fig qnmQ}
\end{figure}

Quasinormal modes are solutions of the wave equation Eqs.  \ref{eqn Vse} and \ref{eqn Vdm1} with boundary conditions of the pure ingoing
waves at the event horizon and pure outgoing waves at infinity or cosmological horizon. In asymptotically anti–de Sitter (AdS) spacetimes,
the boundary at infinity  corresponds to a timelike boundary and requires
different boundary conditions like Dirichlet boundary \cite{I. G Mossand2002}. This makes the study of quasinormal frequencies for AdS black hole incompatible with WKB approximation method. Therefore  the QNMs frequencies of the RNdS-like black hole will be investigated by using the 6th order WKB approximation along with the improvement of worked out 6th order Pade approximation. Refs. \cite{ B.F. Schutz1985,S. Iyer1987} initially proposed to calculate the QNMs frequencies for the non rotating black hole which is also known as WKB method. The extension of WKB method to higher orders has been done in  \cite{konoplya2003,matyjasek2017}. The expression  for finding the QNMs frequencies by applying 6th order WKB method is defined by
\begin{align}
\dfrac{i(\omega^2- V _0)}{\sqrt{-2V''_0}}-\sum_{j=2} ^6 \Omega_j= n+\frac{1}{2}.
\end{align}
where, $V''(r_0)=\dfrac{d^2 V}{dr^2_*} \big | _{r = r_0}, V(r_0)$ represents the peak value of $V(r)_0$ and $r_0$ indicates the value of radial coordinate relating to the maximum of effective potential $V(r)$ and $n$ is the overtone number. It is also noted that  QNMs frequencies $\omega$ can be written in the form $\omega= \omega_R + \omega_I$. The expressions  $\Omega_1, \Omega_2,  \Omega_3, \Omega_4,  \Omega_5$  and  $\Omega_6$ may be found  in \cite{S. Iyer1987,konoplya2003}. If the overtone number $n$ is smaller than multiple number, WKB technique is more credible and accurate. This method is less accurate if $n= \ell$ and is not appropriate for $\ell < n $. To get more accuracy of the higher order WKB method, Refs. \cite{ matyjasek2017,matyjasek2019} proposed the Padé approximation on the usual WKB technique.
Applying the WKB  6th order method and Padé approximation method, the QNMs frequency along with error estimation of RNdS-like black hole are displayed in Tables 1 and 2. For scalar field, electromagnetic field and Dirac field perturbations. We see that increasing the Lorentz violation parameter $L$ and the fixed overtone and multiple numbers, both real and the absolute values of imaginary parts of QNMs frequency increase for the RNdS-like black hole. This shows that increasing the parameter $L$ prevents the rise of both oscillation frequency and damping rate of RNdS-like black hole. In Figs. \ref{fig qnmL} and \ref{fig qnmQ}, we present the variation of real and imaginary parts of the quasinormal frequencies for different values of $L$ and $Q$ corresponding to the fundamental mode for various values of $\ell$, which is obtained by using the Padé approximation.

\begin{longtable}{c c c c c  c  }      
       L & $\ell$ & n &   WKB 6th Order & Padé Approximation & error estimation \\ \hline \multicolumn{6}{c}{SCALAR PERTURBATION} \\ \hline     
       \hline
 0   &	1	&	0	&	0.17902 - 0.0650556$i$ 	&	0.212355 - 0.0755182$i$	&  $1.32465 \times10^{-6}$\\  
    &	2	&	0	&	0.358449 - 0.0731578$i$	&	0.35845 - 0.0731546$i$ 	& $2.55306 \times 10^{-7}$	\\   
    &		&	1	&	0.350628 - 0.220851$i$	&	0.350627 - 0.220851$i$	&	$8.59402 \times10^{-8}$\\   
    &	3	&	0	&	0.503691 - 0.0725371$i$	&	0.503691 - 0.0725367$i$	&	$4.3864\times10^{-8}$\\  
    &		&	1	&	0.497713 - 0.218306$i$	&	0.497713 - 0.218306$i$	& $1.50262\times10^{-7}$	\\  
  0.3  &	1	&	0	&	0.17902 - 0.0650556$i$ 	&	0.179031 - 0.0649739$i$	& $5.53298\times10^{-6}$\\  
    &	2	&	0	&	0.309597 - 0.0635135$i$	&	0.309597 - 0.0635112$i$	&	$2.39344\times10^{-7}$\\ 
    &		&	1	&	0.304006 - 0.191404$i$	&	0.304005 - 0.191398$i$	&	$2.80152\times10^{-6}$\\ 
    &	3	&	0	&	0.437608 - 0.0631554$i$	&	0.437608 - 0.0631552$i$	&	$2.35129\times10^{-8}$	\\  
    &		&	1	&	0.433498 - 0.189854$i$	&	0.433498 - 0.189853$i$	&	$4.65364\times10^{-7}$\\  
  0.6  &			1	&	0	&	0.145429 - 0.0532231 	&	0.145461 - 0.0532054$i$&   0.0000304148	   \\  
    &			2	&	0	&	0.255309 - 0.0524756	$i$&	0.255309 - 0.0524741$i$	&  $1.94131\times10^{-7}$	\\  
    &		&	1	&	0.251892 - 0.157831$i$	&	0.251908 - 0.157825$i$	&	$7.2419\times10^{-6}$	\\ 
    &	3	&	0	&	0.362142 - 0.0523007	$i$&	0.362142 - 0.0523007$i$	&   $1.47197\times10^{-9}$	\\
    &		&	1	&	0.359734 - 0.157064$i$&	0.359732 - 0.157067$i$	&   $1.53396\times10^{-6}$\\ 
        \multicolumn{6}{c}{ELECTROMAGNETIC PERTURBATION} \\ \hline
        
        0   &	1	&	0	&	 0.188803 - 0.0701126$i$ 		&	0.188813 - 0.070056$i$ &	$5.53969\times10^{-6}$\\  
    &	2	&	0	&	0.344942 - 0.0712962	$i$&	0.344942 - 0.071294$i$	&	$2.34059\times10^{-7}$\\   
    &		&	1	&		0.335915 - 0.21522$i$	&	0.335914 - 0.215215 $i$&	$3.02725\times10^{-6}$\\   
    &	3	&	0	&		0.494147 - 0.0716019$i$&	0.494148 - 0.0716016$i$ &	$3.03407\times10^{-8}$\\  
    &		&	1	&		0.487782 - 0.215471$i$	&	0.487782 - 0.21547$i$ &		$4.29148\times10^{-7}$\\  
  0.3  &	1	&	0	&	0.165572 - 0.0615627$i$	&	0.165576 - 0.0615174$i$ & $1.69517\times10^{-6}$	\\  
    &	2	&	0	&		0.301784 - 0.0623734	$i$&	0.301784 - 0.0623719$i$&  $3.49468\times10^{-8}$	\\ 
    &		&	1	&	0.295795 - 0.187773$i$	&	0.295795 - 0.187773 $i$&	$8.20042\times10^{-8}$\\ 
    &	3	&	0	&		0.431932 - 0.190376$i$	&  0.431932 - 0.190375$i$ &	$4.41179\times10^{-9}$\\  
    &		&	1	&		0.427845 - 0.188091$i$	&	0.427845 - 0.188091$i$ & $1.09204\times10^{-8}$	\\  
  0.6  &			1	&	0	&		0.137952 - 0.0513998$i$&  0.137949 - 0.0513782$i$ &   $1.33016\times10^{-6}$\\  
    &			2	&	0	&		0.250881 - 0.0518711	$i$&  0.250881 - 0.0518695$i$ &	 $6.47132\times10^{-8}$\\  
    &		&	1	&		0.247487 - 0.15586$i$	&  0.247488 - 0.15586$i$ &	$2.43434\times10^{-6}$\\ 
    &	3	&	0	&		0.358995 - 0.0519975	$i$ &	0.358995 - 0.0519973$i$ &	$8.18921\times10^{-9}$\\
    &		&	1	&	 0.356591 - 0.15612$i$ 		&	0.356591 - 0.15612$i$ &		$3.6078\times10^{-7}$\\ 
         \multicolumn{6}{c}{DIRAC PERTURBATION} \\ \hline
        
  0  &	1	&	0	&	0.302038 - 0.0739107$i$	&  0.301757 - 0.0739997$i$	&		$5.14345\times10^{-6}$	\\  
    &	2	&	0	&	0.454395 - 0.0740455$i$		&	0.454402 - 0.074044$i$	&	$1.16507\times10^{-6}	$		\\ 
    &		&	1	&	0.446769 - 0.223158$i$	&	0.446863 - 0.223099$i$	&		$	0.0000134907$	\\
    &	3	&	0	&	0.60669 - 0.0740631$i$	&	0.606691 - 0.074063$i$	&	$5.08025\times10^{-7}	$		\\    
    &		&	1	&	0.601025 - 0.222737$i$	&	0.601034 - 0.222729$i$	&	$4.11294\times10^{-6}	$		\\  
 0.3   &		1&	0	&	0.267388 - 0.0654677	$i$&	0.267044 - 0.065557$i$	&		$9.27225\times10^{-6}	$	\\
    &	2	&	0	&	0.404581 - 0.0658356$i$	&	0.404573 - 0.0658364$i$&	$2.31002\times10^{-6}$		\\    
    &		&	1	&	0.399347 - 0.198003	$i$&	0.399352 - 0.198047$i$	&		$0.0000963835	$	\\  
    &	3	&	0	&	0.54133 - 0.0659304$i$	&	0.54133 - 0.0659301$i$	&		$7.5831\times10^{-7}	$	\\ 
    &		&	1	&	0.537352 - 0.198092$i$	&	0.537353 - 0.19809$i$	&		$	4.26594\times10^{-6}$	\\
 0.6   &	1	&	0	&	0.230047 - 0.0563368$i$	&	0.229635 - 0.0564307$i$&		$0.0000128137$	\\  
    &	2	&	0	&	0.348641 - 0.056722	$i$&	0.348808 - 0.0566936$i$	&		$2.7598\times10^{-6}	$	\\ 
    &		&	1	&	0.343608 - 0.171247$i$	&	0.34544 - 0.17033$i$ 	&		0.000011877	\\
    &	3	&	0	&	0.467113 - 0.0567813$i$	&	0.467126 - 0.05678$i$	&		$8.90456\times10^{-7}	$	\\ 
    &		&	1	&	0.464458 - 0.170531$i$	&	0.464607 - 0.17048$i$	&		$3.51822\times10^{-6}$	\\  
    \caption{QNM frequencies for all perturbations of RNdS-like black hole derived by using 6th order WKB and 6th order Padé Approximation for different modes and for different values of the charge $L$ with fixed $M=1$, $\Lambda=0.05 $ and $Q=0.2.$}
    \label{tab 	Qnms 1}
\end{longtable}

\begin{longtable}{c c c c c  c  }      
       Q & $\ell$ & n &   WKB 6th Order & Padé Approximation & error estimation \\ \hline \multicolumn{6}{c}{SCALAR PERTURBATION} \\ \hline     
        0   &	1	&	0	&	0.182786 - 0.0692193	$i$&	0.182778 - 0.0691354$i$	&		$2.21674\times10^{-6}$	\\  
    &	2	&	0	&	0.318619 - 0.0665137$i$&	0.31862 - 0.0665106$i$	&		$4.17175\times10^{-7}$	\\ 
    &		&	1	&	0.312615 - 0.200609$i$	&	0.312613 - 0.200606$i$	&		$1.222\times10^{-6}$	\\
    &	3	&	0	&	0.451375 - 0.0658817$i$&	0.451375 - 0.0658814$i$&		$3.65515\times10^{-8}$	\\ 
    &		&	1	&	0.446795 - 0.198135$i$&	0.446795 - 0.198134$i$	&		$3.01163\times10^{-7}	$	\\  
    &		&	2	&	0.437808 - 0.331875$i$	&	0.437808 - 0.331874$i$&		$8.51647\times10^{-7}	$	\\ \hline
    0.3   &	1	&	0	&	0.190118 - 0.0708744$i$	&0.190107 - 0.0707981$i$		&		$2.50536\times10^{-6}$	\\  
    &	2	&	0	&	0.330613 - 0.0681468$i$&	0.330613 - 0.0681437$i$	&		$3.84439\times10^{-7}$	\\ 
    &		&	1	&	0.324436 - 0.205538$i$	&0.324435 - 0.205536	$i$ & 	$9.76281\times10^{-7}$		\\
    &	3	&	0	&	0.468112 - 0.0675029$i$&	0.468112 - 0.0675026$i$&		$4.62965\times 10^{-8}$	\\ 
    &		&	1	&	0.46338 - 0.203023$i$ 	&	0.46338 - 0.203022$i$	&		$4.85972\times10^{-7}$	\\  
    &		&	2	&	0.454108 - 0.340103$i$	&	0.454107 - 0.340102$i$	&		$8.69836\times10^{-7}$	\\ \hline
    0.6   &	1	&	0	&	0.215192 - 0.0756086$i$&	0.215184 - 0.0755466$i$	&		$1.50508\times10^{-6}$	\\  
    &	2	&	0	&	0.371494 - 0.0729003$i$&	0.371494 - 0.0728971$i$	&		$5.89372\times10^{-7}$	\\ 
    &		&	1	&	0.364936 - 0.21986$i$ 	&	0.364934 - 0.219856$i$	&	$1.29278\times10^{-6}$		\\
    &	3	&	0	&	0.525089 - 0.072241$i$	&	0.525089 - 0.0722406$i$	&		$2.83432\times10^{-8}	$	\\ 
    &		&	1	&	0.520019 - 0.217299$i$	&	0.520019 - 0.217299$i$	&		$1.18359\times10^{-7}	$	\\  
    &		&	2	&	0.510122 - 0.364088$i$	&	0.510122 - 0.364088$i$	&		$6.4087\times10^{-7}$	\\  \hline \hline
    \multicolumn{6}{c}{ELECTROMAGNETIC PERTURBATION} \\ \hline     
        0   &	1	&	0	&	0.17087 - 0.0638571$i$	&0.170875 - 0.0638125$i$ 		&		$1.55934\times10^{-6}	$	\\  
    &	2	&	0	&	0.311815 - 0.0647803$i$	&	0.311815 - 0.0647786$i$	&		$6.11989\times10^{-8}	$	\\ 
    &		&	1	&	0.304974 - 0.195158$i$	&	0.304974 - 0.195158$i$	&	$4.66599\times10^{-7}$		\\
    &	3	&	0	&	0.446564 - 0.0650204$i$&	0.446564 - 0.0650202$i$	&		$7.76202\times10^{-9}$	\\ 
    &		&	1	&	0.441738 - 0.195474$i$&	0.441738 - 0.195474$i$	&		$6.39638\times10^{-8}$	\\  
    &		&	2	&	0.432043 - 0.327267$i$	&	0.432058 - 0.327258$i$	&		$0.0000197843$	\\ \hline
    0.3   &	1	&	0	&	0.177379 - 0.0654266	$i$&	0.177384 - 0.0653753$i$&		$2.96882\times10^{-6}$	\\  
    &	2	&	0	&	0.323358 - 0.0663726$i$	&	0.323358 - 0.0663707$i$	&		$1.03232\times10^{-7}	$	\\ 
    &		&	1	&	0.316258 - 0.200003$i$&	0.316258 - 0.200001$i$	&	$2.12005\times10^{-6}$		\\
    &	3	&	0	&	0.462986 - 0.0666204$i$&	0.462986 - 0.0666202$i$	&		$1.31765\times10^{-8}$	\\ 
    &		&	1	&	0.457978 - 0.200307$i$	&	0.457978 - 0.200307$i$&		$3.042\times10^{-7}$	\\  
    &		&	2	&	0.447936 - 0.335434$i$&	0.447951 - 0.33542$i$&		$0.00002152$	\\ \hline
    0.6   &	1	&	0	&	0.199825 - 0.0700572$i$&0.199829 - 0.0699976$i$		&		$1.20803\times10^{-6}	$	\\  
    &	2	&	0	&	0.362769 - 0.0710499$i$&	0.362769 - 0.0710476$i$&		$4.76638\times10^{-8}$	\\ 
    &		&	1	&	0.355081 - 0.214222$i$&	0.35508 - 0.21422$i$	&	$1.64759\times10^{-6}$		\\
    &	3	&	0	&	0.518933 - 0.071317$i$	&	0.518933 - 0.0713167$i$	&		$	6.05469\times10^{-9}$	\\ 
    &		&	1	&	0.513502 - 0.21449$i$	&	0.513502 - 0.21449$i$	&		$2.19292\times10^{-7}$	\\  
    &		&	2	&	0.502695 - 0.359377$i$	&	0.502695 - 0.359367$i$&		$0.0000104945$	\\ \hline \hline
    \multicolumn{6}{c}{DIRAC PERTURBATION} \\ \hline     
        0   &	1	&	0	&	0.256971 - 0.064847$i$ 	&	0.256969 - 0.0648489$i$	&		$3.08449\times10^{-6}$	\\  
    &	2	&	0	&	0.38883 - 0.0650885$i$&	0.388833 - 0.0650886$i$ &		$9.65292\times10^{-7}$	\\ 
    &		&	1	&	0.383222 - 0.19584$i$&	0.383246 - 0.195826$i$	&		$0.0000124913$	\\
    &	3	&	0	&	0.520048 - 0.0651693$i$&	0.520047 - 0.0651695$i$&		$2.04727\times10^{-7}$	\\ 
    &		&	1	&	0.515868 - 0.195823$i$&	0.515865 - 0.195824	$i$&		$3.20641\times10^{-6}$	\\  
    &		&	2	&	0.507508 - 0.327464$i$	&	0.507496 - 0.327466$i$&		$	0.0000145777$	\\ \hline
    0.3   &	1	&	0	&	0.291249 - 0.0700287$i$&	0.290806 - 0.0701469$i$&		$7.29503\times10^{-6}$	\\  
    &	2	&	0	&	0.439859 - 0.0703906$i$	&	0.439871 - 0.0703881$i$&		$	1.94222\times10^{-6}$	\\ 
    &		&	1	&	0.433527 - 0.211944$i$	&	0.433664 - 0.211871$i$&		$   0.0000139827$	\\
    &	3	&	0	&	0.588228 - 0.0704712$i$	&	0.588223 - 0.0704718$i$&		$	1.24609\times10^{-6}$	\\ 
    &		&	1	&	0.583652 - 0.211786$i$	&	0.583594 - 0.21181$i$&		$0.0000156872	$	\\  
    &		&	2	&	0.574722 - 0.354152$i$	&	0.574326 - 0.354344$i$&		$0.0000366864$	\\ \hline
    0.6   &	1	&	0	&	0.335319 - 0.0746014$i$&	0.3347 - 0.0747522$i$&		$5.67511\times10^{-6}$	\\  
    &	2	&	0	&	0.506072 - 0.0749514$i$&	0.505841 - 0.0749843$i$&		$1.95925\times10^{-6}$	\\ 
    &		&	1	&	0.502034 - 0.224583$i$&	0.499436 - 0.22573$i$&	$0.0000119615$		\\
    &	3	&	0	&	0.676236 - 0.0750675$i$	&	0.676237 - 0.0750666$i$	&		$7.36215\times10^{-7}	$	\\ 
    &		&	1	&	0.671429 - 0.225649$i$&	0.671435 - 0.225636$i$	 &		$	3.84269\times10^{-6}$	\\  
    &		&	2	&	0.661887 - 0.377623$i$	&	0.66188 - 0.377512$i$	&		$0.0000495886$	\\ 
   
    \caption{QNM frequencies for all perturbations of RNdS-like black hole derived by using 6th order WKB and 6th order Padé Approximation for different modes and for different values of the charge $Q$ with fixed $M=1$, $\Lambda=0.05 $ and $L=0.2.$}
    \label{tab qnms 2}
\end{longtable}.

\section{Evolution of scalar, electromagnetic and Dirac perturbations}\label{sec 8}
In this section, we will study the time domain profiles of scalar, electromagnetic and Dirac field perturbations. This will help in a better understanding of black hole dynamical response under these perturbations. To obtain the time evolution, we numerically solve the time-dependent wave equation using the time domain integration formalism introduced by Gundlach et al. \cite{C. Gundlach1994}.  We define  the wave function as 
$\psi(r_*, t)$, which is discretized on a numerical grid as $\psi(i \Delta r_{*},j \Delta t) = \psi_{i,j}$. Similarly, the effective potential is expressed as 
\(V(r(r_*)) = V(r_*, t) = V_{i,j}\). Now, the governing  equation   can be expressed in the form 

\begin{align}
\frac{\psi_{i+1,j} - 2\psi_{i,j} + \psi_{i-1,j}}{(\Delta r_*)^2}
- \frac{\psi_{i,j+1} - 2\psi_{i,j} + \psi_{i,j-1}}{(\Delta t)^2}
- V_i \psi_{i,j} = 0.
\end{align}
We use the initial condition, $\psi(r_*, t) = \exp\left[ -\frac{(r_* - \tilde{k}_1)^2}{2\sigma^2} \right]$
and $\psi(r^*, t) \big|_{t < 0} = 0$,
where $\tilde{k}_1$ and $\sigma$ represent the median and width of the initial wave packet respectively. Now applying the iterative scheme,  the time evolution of the scalar, electromagnetic and Dirac field perturbations can be calculated as
\begin{align}
\psi_{i,j+1} = -\psi_{i,j-1} 
+ \left( \frac{\Delta t}{\Delta r_*} \right)^2 \left(\psi_{i+1,j} + \psi_{i-1,j}\right) 
+ \left[ 2 - 2 \left( \frac{\Delta t}{\Delta r_*} \right)^2 - V_i \Delta t^2 \right] \psi_{i,j}.
\end{align}
Utilizing the above iteration scheme, one can obtain the profile of $\psi$ with respect to time $t$ by choosing the appropriate values of $\Delta t$ and $\Delta r_*$ such that it satisfies the Von Newman stability condition $\frac{\Delta t}{\Delta r_*} <1$.

\begin{figure}[h!]
\centering
  \subfloat[\centering ]{{\includegraphics[width=140pt,height=140pt]{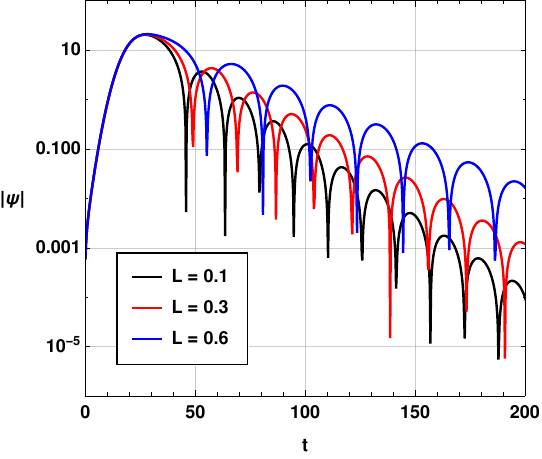}}\label{fig scalartdL}}
  \quad
   \subfloat[\centering ]{{\includegraphics[width=140pt,height=140pt]{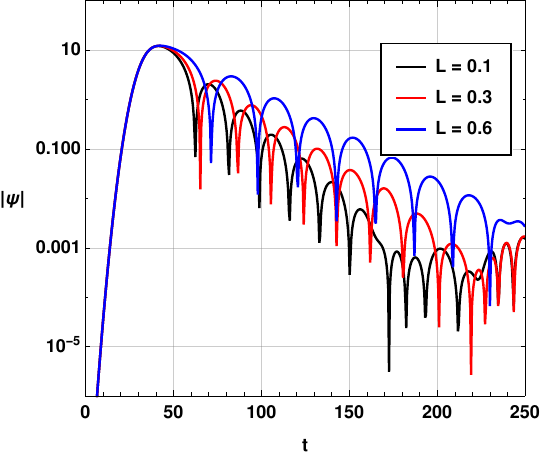}}\label{fig EMtdL}}
   \quad
   \subfloat[\centering ]{{\includegraphics[width=140pt,height=140pt]{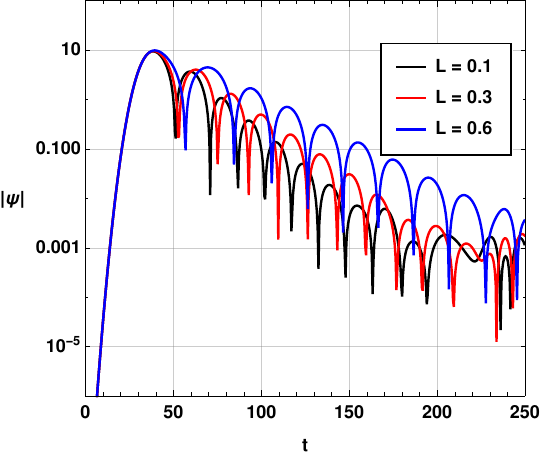}}\label{fig DiractdL}}
   \caption{Time domain profile of (a) scalar field perturbation, (b) electromagnetic field perturbation and (c) Dirac field perturbation for RNdS-like black hole for different values of $L$. The parameters are taken as  $L=0.2$, $M=1$, $l=1$ and $\Lambda=0.05$. }
   \label{fig tdL}
\end{figure}



\begin{figure}[h!]
\centering
  \subfloat[\centering ]{{\includegraphics[width=140pt,height=140pt]{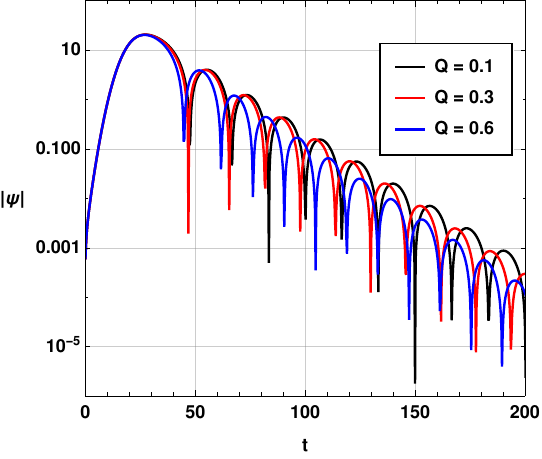}}\label{fig scalartdL}}
  \quad
   \subfloat[\centering ]{{\includegraphics[width=140pt,height=140pt]{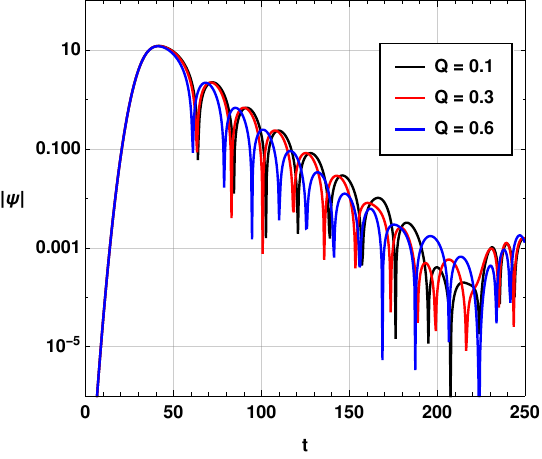}}\label{fig EMtdL}}
   \quad
   \subfloat[\centering ]{{\includegraphics[width=140pt,height=140pt]{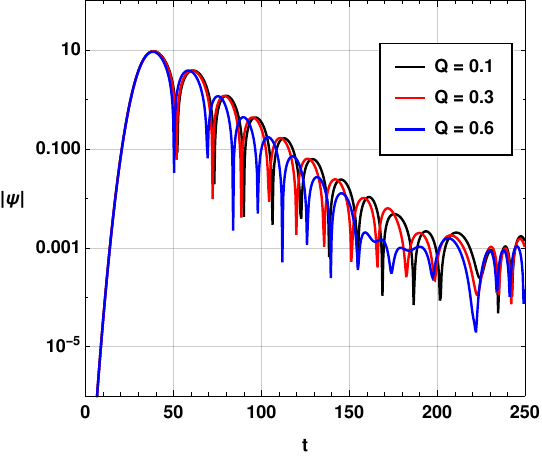}}\label{fig DiractdL}}
   \caption{Time domain profile of (a) scalar field perturbation, (b) electromagnetic field perturbation and (c) Dirac field perturbation for RNdS-like black hole for different values of $Q$. The parameters are taken as  $L=0.2$, $M=1$, $l=1$ and $\Lambda=0.05$. }
   \label{fig tdQ}
\end{figure}

In Fig. \ref{fig tdL}, we illustrate the impact of $L$ on the time domain profile of scalar, electromagnetic and Dirac perturbations for  RNdS-like  black hole. We observe a significant distinction in the time-domain profiles corresponding to different values of $L$ in all  types of perturbations. Increasing the values of $L$ result in a reduction of the oscillation frequency across all the perturbations. Further, with increasing $L$, the magnitude of the slopes of the maxima in the logarithmic graph decrease, indicating a slower damping rate. 
Figure \ref{fig tdQ} illustrates the influence of $Q$ on the time domain evolution of scalar, electromagnetic and Dirac perturbations for  RNdS-like  black hole. In all the three perturbations, the damping rate  and oscillation frequencies are enhanced with increasing $Q$. 

\section{Null geodesic and shadow radius of black hole}\label{sec 9}
The study of photon sphere and black hole shadow becomes an important topic in the study of black holes within the framework of bumblebee gravity. The modification of spacetime geometry in bumblebee field can alter the the radius of the photon sphere, leading to observable differences in the shadow cast by the black hole. In the context of bumblebee gravity, Refs. \cite{A. Uniyal2023,izmailov2022,karmaker2023} studied the photon radius and black hole shadow radius of different black holes.  
 
\begin{figure}[h!]
\centering
  \subfloat[\centering ]{{\includegraphics[width=200pt,height=170pt]{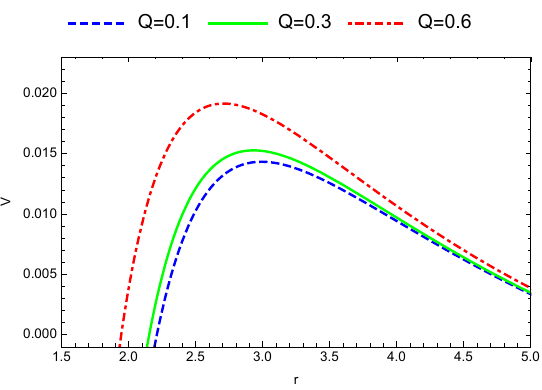}}\label{fig 16a}}
  \qquad
   \subfloat[\centering ]{{\includegraphics[width=200pt,height=170pt]{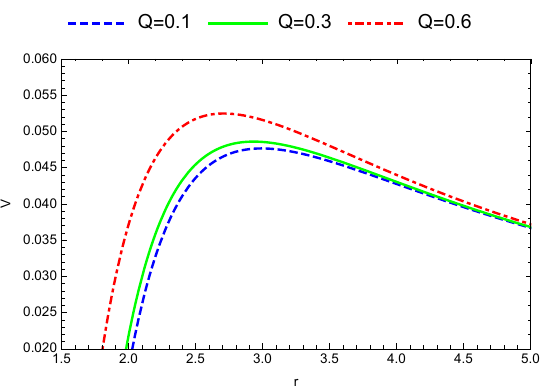}}\label{fig 16b}}
   \caption{Plot of effective potential for null geodesic for different values of charge $Q$ (a) with fixed $L=0.2$, $M=1$ and $\Lambda=0.05$. (b) with fixed $L=0.2$, $M=1$ and $\Lambda=-0.05$.}
   \label{fig 16}
\end{figure}

\begin{figure}[h!]
\centering
  \subfloat[\centering ]{{\includegraphics[width=170pt,height=170pt]{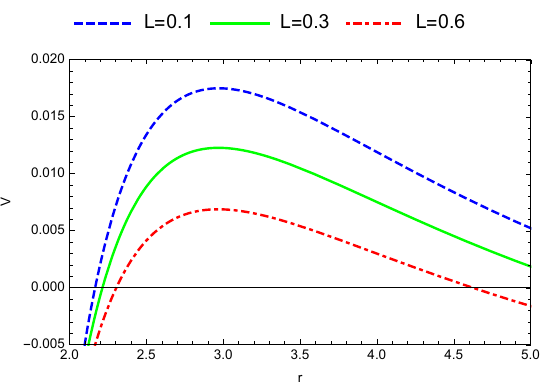}}\label{fig 17b}}
  \qquad
   \subfloat[\centering ]{{\includegraphics[width=170pt,height=170pt]{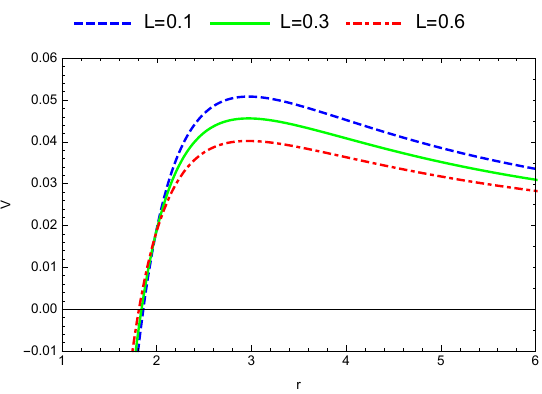}}\label{fig 17b}}
   \caption{ Plot of effective potential for null geodesic for different values of charge $L$ (a) with fixed $Q=0.2$, $M=1$ and $\Lambda=0.05$. (b) with fixed $Q=0.2$, $M=1$ and $\Lambda=-0.05$.}
   \label{fig 17}
\end{figure}

 In this section the photon's orbit and radius of the RNdS-like black hole will be investigated. The Lagrangian $ \mathscr{L}(x,\dot{x})=\frac{1}{2}g_{ab} \dot{x}^a\dot{x}^b$ for the RNdS-like black hole is defined by
\begin{align}\label{eqn lag1}
2   \mathscr{L}= A(r) \dot{t}^2 -\dfrac{(1+L) \dot{r}^2}{A(r)}- r^2 \dot{\theta}^2 - r^2 \sin^2\theta \dot{\varphi}^2,
\end{align}
where the dot indicates the differentiation with respect to an affine parameter $\tau$. Since the above equation is static and spherically symmetric spacetime, the energy $E=p_a\xi^a_{(t)}$ and the angular momentum $\Lb=-p_a\xi^a_{(\phi)}$ along the geodesic will be conserved. $\xi_{t}^a$ and $\xi_{t}^a$ denote the Killing vectors due to time translational and rotational invariance. Hence $E=p_t$ and $\Lb=-p_{\phi}$ are the energy of a photon and the angular momentum respectively. Since $p_\alpha= \dfrac{d \Lb}{d\dot{\alpha}}$, where $p_{\alpha}$ represents the conjugate momentum corresponding to the coordinate ${\alpha}$,
we derive the following equations in the equilateral plane
\begin{align}\label{eqn lag2}
p_t=A(r) \dot{t}=E, \hspace{1cm} p_r=\dfrac{1+L}{A(r)} \dot{r}, \hspace{1cm} p_\phi=-r^2\dot{\phi}=-\Lb.
\end{align}
Using Eq. (\ref{eqn lag2}) in Eq. (\ref{eqn lag1}), the differential equation for null geodesic is calculated as  
\begin{align}
\dot{r}^2+V(r)=0,
\end{align}
 where V(r) denotes the potential which is defined by
\begin{align}\label{eqn V}
V(r)=\dfrac{1}{1+L} \left(\dfrac{A(r)\Lb}{r^2}-E^2 \right).
\end{align}
The graphs of effective potential for both RNdS and RNAdS-like black holes with respect to $r$ are drawn in Figs. \ref{fig 16} and \ref{fig 17} with varying $Q$ and $L$ respectively. Increasing $Q$ increases the peak of effective potential for both  RNdS and RNAdS-like black holes. However increasing the parameter $L$ decreases the peak of effective  potential for both the black holes. We also see that the peak of effective potential shift towards the left in all cases. There exists the unstable circular orbits located at the peak of the above potential.


\begin{figure}[h!]
\centering
  \subfloat[\centering ]{{\includegraphics[width=170pt,height=170pt]{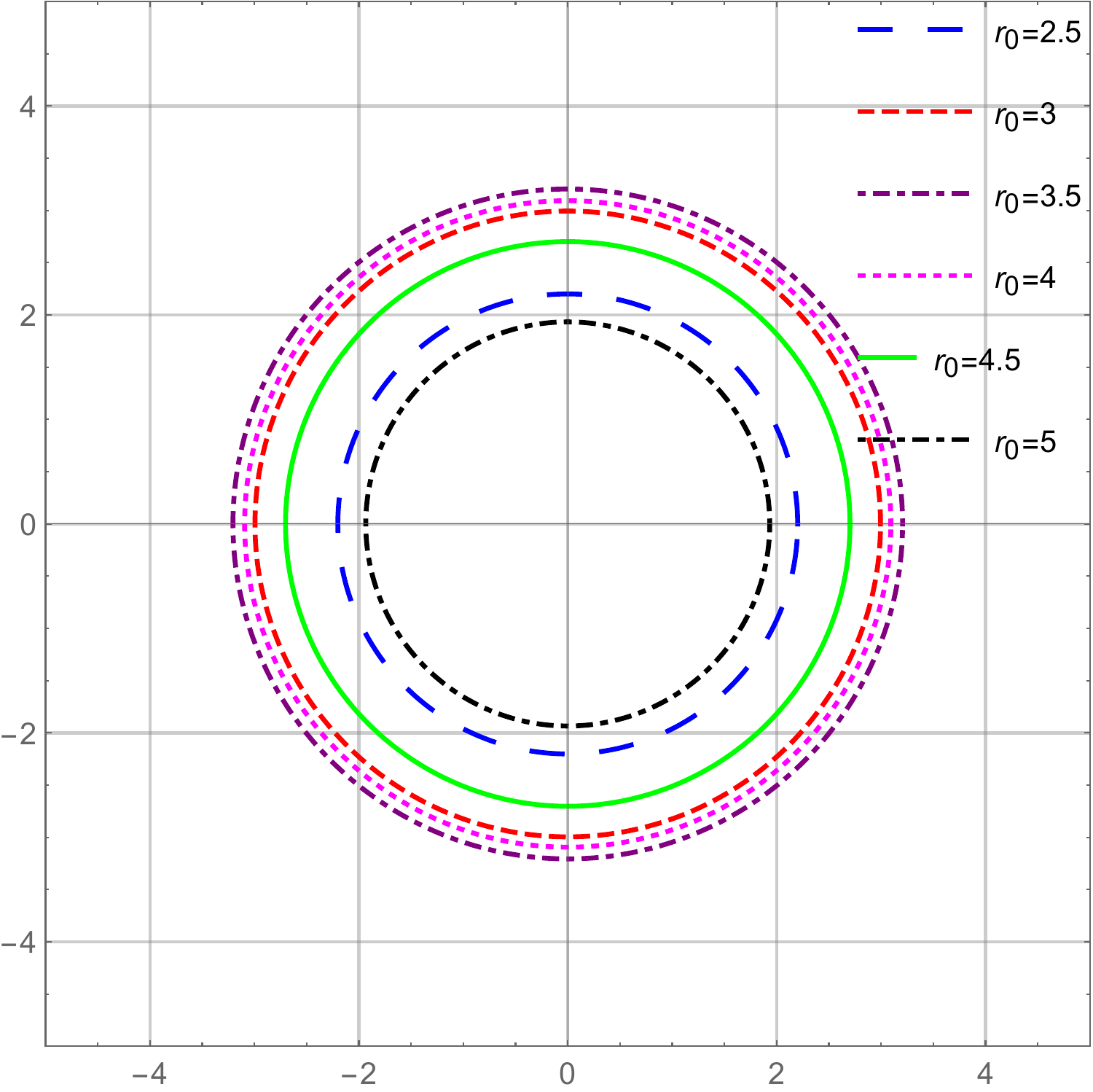}}\label{fig dSr0}}
  \qquad
   \subfloat[\centering ]{{\includegraphics[width=170pt,height=170pt]{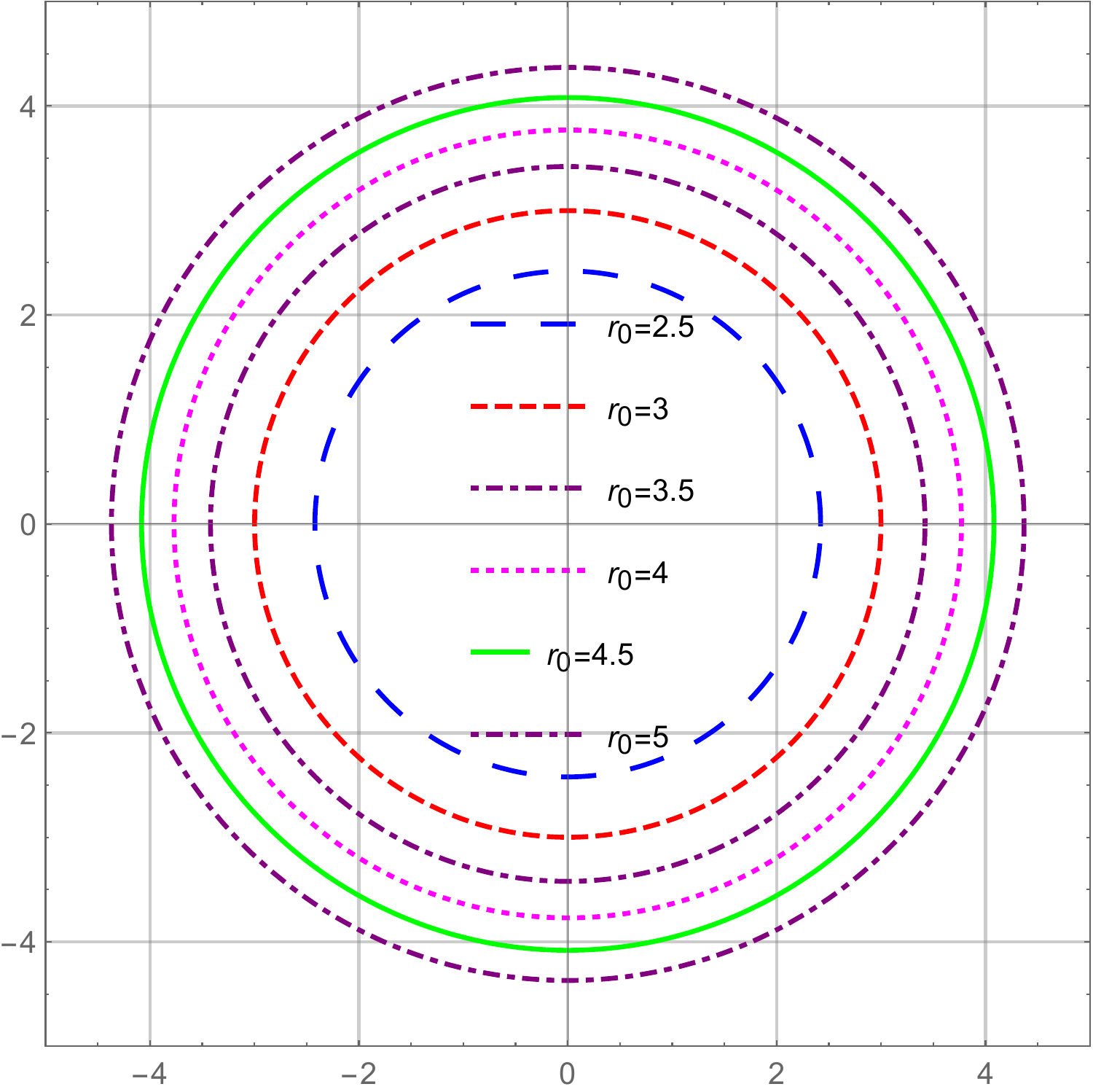}}\label{fig AdSr0}}
   \caption{Plot of the shadow radius varying $r_0$ for (a) RNdS-like black hole ($\Lambda= 0.05$)  (b) RNAdS-like black hole ($\Lambda=-0.05)$    with fixed $L=0.3$, $Q=0.3$ and  $M=1$.}
   \label{fig shadowr0}
\end{figure}

\begin{figure}[h!]
\centering
  \subfloat[\centering ]{{\includegraphics[width=170pt,height=170pt]{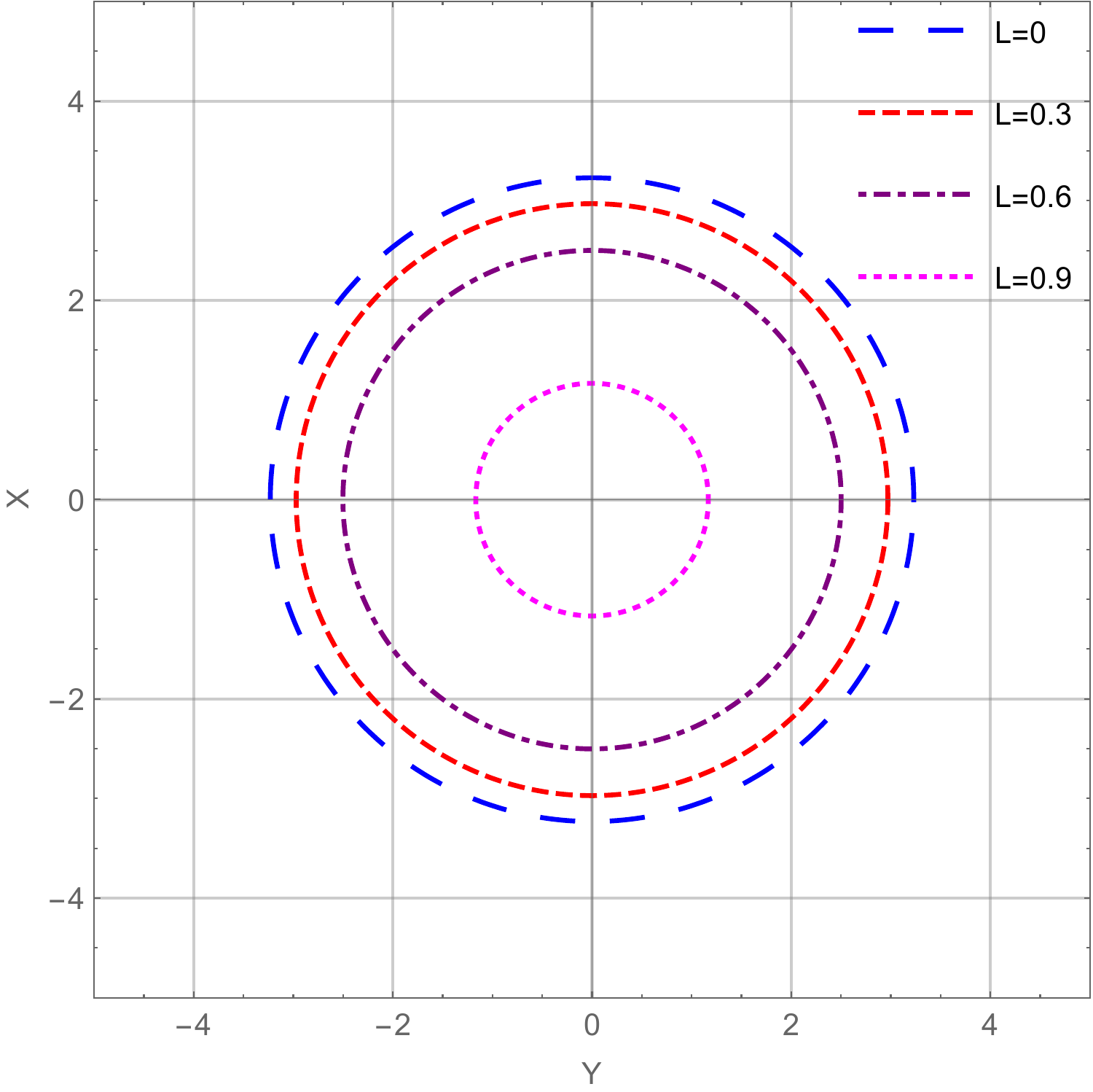}}\label{fig shadowLdS}}
  \qquad
   \subfloat[\centering ]{{\includegraphics[width=170pt,height=170pt]{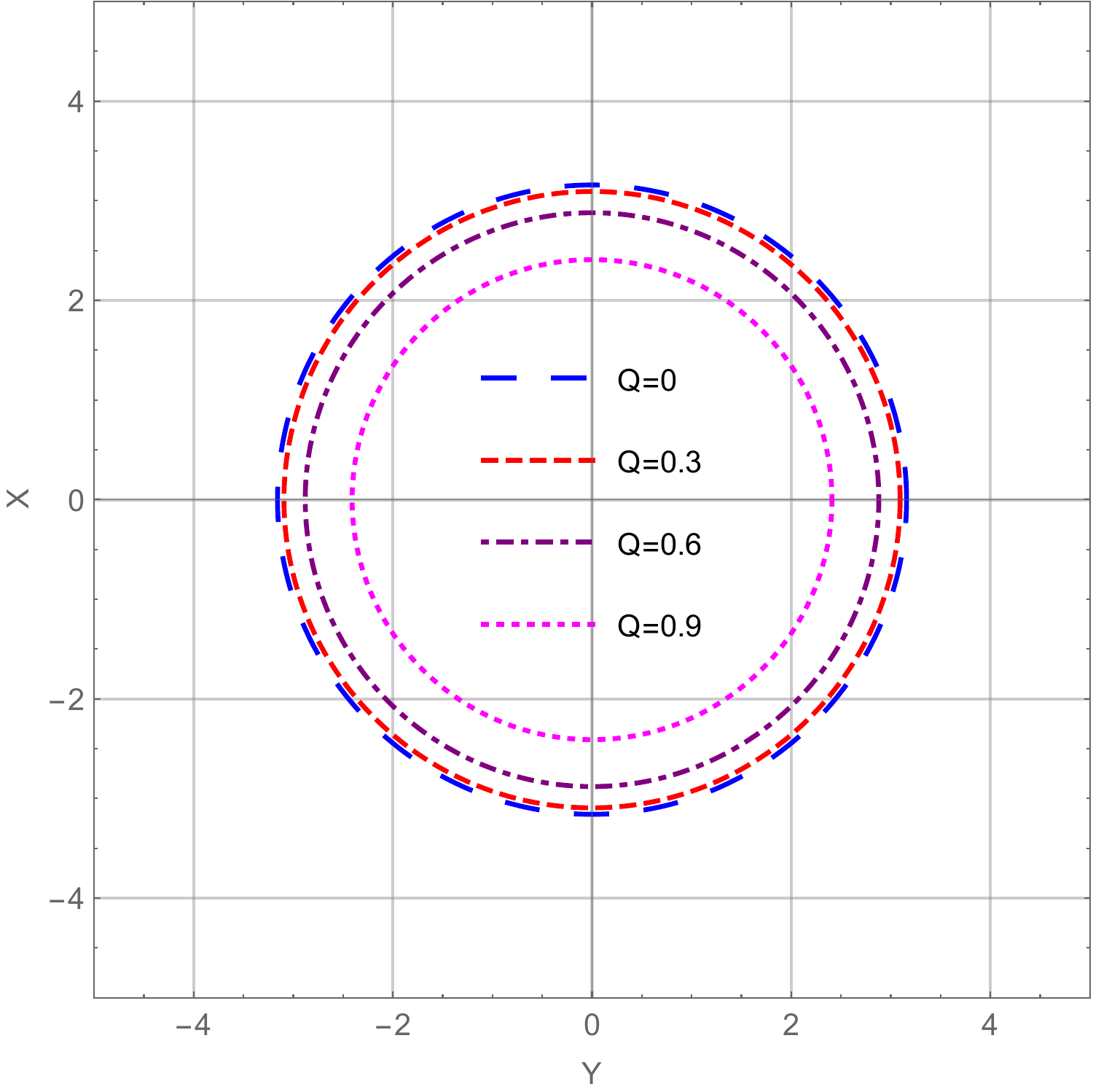}}\label{fig shadowQdS}}
   \caption{ Plot of the shadow radius of  RNdS-like black hole  for different values of  $L$ and $Q$. The set of parameters are taken as (a) $r_0=4$, $\Lambda=0.05$, $M=1$ and $Q=0.5$, (b) $r_0=4$, $\Lambda=0.05$, $M=1$ and $L=0.3$. }
   \label{fig shadow}
\end{figure}

The circular photon orbits of radius $r_p$ of RNdS-like black hole should satisfy the following conditions
 \begin{align}\label{eqn Vrp1}
V(r)\vert_{r=r_p}=0, \hspace{0.5cm} \hspace{0.5cm}  V'(r)\vert_{r=r_p}=0 \hspace{0.5cm} \rm and \hspace{0.5cm}   V''(r)\vert_{r=r_p}=0.
\end{align}
On simplification of the middle equation of Eq. \eqref{eqn Vrp1}, we get
\begin{align}\label{eqn Vrp2}
 2A(r_{p})-r_{p}A'(r_{p})=0,
\end{align}
where $r_p$ represents the photon sphere radius at $r=r_p$.
One may also define the critical impact parameter for the photon sphere radius as 
\begin{align}
b_c=\dfrac{\Lb}{E}=\dfrac{r_p}{\sqrt{A (r_p)}}.
\end{align}
 We derive the photon sphere radius from Eq. \eqref{eqn Vrp2} as
 \begin{align}\label{eqn VrpM}
r_p=\dfrac{3 M}{2}+ \dfrac{1}{2}\sqrt{9M^2-\dfrac{16(1+L)Q^2}{2+L}}.
\end{align}
From above equation, the photon sphere radius depends not only on mass of the black hole $M$, charge $Q$ but also on $L$. It is important to note that for Schwarzschild and Schwarzschild-de Sitter spacetime in bumblebee gravity, the photon radius is unaffected by $L$ \cite{maluf2021,Y.S.Priyobarta2024,izmailov2022}. It is also noted that the real photon radius is obtained if $L< (18M^2- 16Q^2)/( 16Q^2-9M^2)$ and when $Q=0$, 	Eq. (54) reduces to photon sphere radius of Schwarzchild black hole.  Increasing the mass of the black hole increases the photon sphere radius for both RNdS and RNAds-like black hole. However the photon sphere radius decreases with the increase of $Q$ and $L$ as calculated in the Table 3. For a static observer located at $r=r_0$, the black hole shadow radius $R_s$ is given by \cite{konoplya2019b}
 
\begin{align}\label{eqn Rs1}
R_s= {\dfrac{r_p \sqrt{A(r_0)}}{\sqrt{A(r_p)}}}.
\end{align}
From Eqs. \eqref{eqn VrpM} and \eqref{eqn Rs1}, it is noted that the photon sphere radius is independent of cosmological constant $\Lambda$, but the shadow radius depends on $\Lambda$. If the observer is located at large distance ($r_o\rightarrow\infty$) and $A(r_0)\rightarrow 1$, the spacetime becomes asymptotically flat space. In such case, the black hole shadow radius can be written as
\begin{align}
R_s= \dfrac{r_p}{\sqrt{A(r_p)}}.
\end{align}
 Thus the apparent form of the black hole shadow is independent of the position of observers for asymptotically flat black hole with a static observer located at large distance. However for non-asymptotically flat black hole, the appearance of the black hole shadow radius is affected by the position of the observer $r_0$ \cite{perlick2022}. 
The shadow radius can also be represented by the celestial coordinate $(X,Y)$ as
\begin{align}
 &X= \lim\limits_{r_0 \to \infty} \left(-r_{0}^2  \sin\theta_0 \frac{d\phi}{dr}\vert_{r_0,\theta_0} \right),   ~~~Y= \lim\limits_{r_0 \to \infty} \left(r_{0}^2  \frac{d\theta}{dr}\vert_{r_0,\theta_0} \right), 
\end{align}
where $\theta_0$ is the position of the observer along the plane of black hole.  We shall apply Eq. \eqref{eqn Rs1} to derive the stereographic projections of RNdS and RNAdS-like black holes shadows where the physical observer is located at $r=r_0$. It is noted that for de Sitter black hole, the observer must lie between event and cosmological horizons i.e. $r_h<r_0<r_c$.
For fixed $Q$, $L$, $\Lambda$ and $M$,  it is observed from Eq. \eqref{eqn Rs1} that $R_s \propto  A({r}_0)$. The graph of $A(r_{0})$ for RNdS-like black hole increases monotonically if $r_h<r_0< r_{max}$ and decreases monotonically if $r_{max}<r_0<r_c$, where $r_{max}$ is the location of maximum $A(r_0)$ as shown in Figs.  \ref{fig 1a}  and \ref{fig 2a}. Therefore the radius of black hole shadow of RNdS-like black hole increases if $r_h<r_0\leq r_{max}$ and decreases if $r_{max}<r_0<r_c$ as shown in Fig. \ref{fig dSr0}. For Fig. \ref{fig dSr0},  the set of parameters are taken as $M=1$, $\Lambda=0.05$, $Q=0.3$ and $L=0.3$. Here $r_{max}=3.55246$ and the physical observer distance lies between $r_h=2.17672$ and $r_c=5.4079$. However, for the RNAdS-like black hole, $A(r_0)$ increases monotonically if $r_h<r_0<\infty$ as shown in Figs.  \ref{fig 1b}  and \ref{fig 2b}. 
Hence with increasing the distance of the observer from the event horizon, the observer will see a larger black hole shadow for RNAdS-like black hole  as shown in Fig. \ref{fig AdSr0}. 
To investigate the impact of $L$ and $Q$ on the black hole shadow radius, we draw the stereographic projection of the RNdS-like black hole shadow where a finite observer is located at $r_0=4$ $(r_h<r_0<r_c)$ in Fig. \ref{fig shadow}.  We see from Fig. \ref{fig shadow} that increasing $L$ and $Q$ prevent the rise of RNdS-like black hole shadow radius. If $L=0$ and 
$Q=0$ in Fig. \ref{fig shadow}, these correspond to the shadow radius of RNdS and SdS-like black holes respectively. It is observed that the shadow radius of the RNdS-like black hole is smaller than that of the RNdS and SdS-like black hole. 



\begin{table}[] \label{tab 3}
\begin{tabular}{c| c c c| c c c| c c c}
 $L$& \multicolumn{3}{|c}{0.1}  & \multicolumn{3}{|c}{0.3}  &  \multicolumn{3}{|c}{0.6} \\ \hline
 $Q$&  0.1 & 0.3 & 0.6 & 0.1 & 0.3 & 0.6 & 0.1 &0.3  &0.6  \\ \hline
 $r_p$& 2.993 &  2.93577 & 2.72299 & 2.99244 & 2.93057 & 2.69837 & 2.99177 & 2.92424 & 2.66784 \\ \hline
 $R_{s+}$ &	1.5107	&1.50607	& 1.47589	& 1.23685	& 1.23624	& 1.21764	& 0.722371	& 0.738528	&  0.76916  \\ \hline
 $ R_{s-}$ &	3.37028	& 3.35982	&	3.30748&  3.4932	&3.48189	& 3.42361	& 3.67013	& 3.65783	& 3.59216
\end{tabular}
\caption{Values for photon radius and black hole shadow radius of RNdS ($R_{s+}$) and RNAdS-like ($R_{s-}$) black hole for various values of $L$ and Q. Here we set $M=1$, $\Lambda=0.05(-0.05)$ and $r_0=4$.}
\end{table}

\section{Discussions and Conclusion}\label{sec 11}
In this paper, the Dirac, scalar and electromagnetic perturbations are studied within the frame work of RNdS and RNAdS-like black holes in bumblebee gravity. The effective potentials of RNdS and RNAdS-like black holes for all perturbations and the impacts of $Q$ and $L$ in the effective potentials, greybody factors, QNMs are investigated. We see that the behaviour of effective potentials remain the same  in the region between cosmological and event horizons for all the perturbations. For both RNdS and RNAdS-like black holes,  the height of effective potentials increase with the increase of $Q$  in all the perturbations.  For scalar  and electromagnetic field perturbations, increasing $L$ decreases the effective potential for RNdS-like black hole but it has opposite effect in RNAdS-like black hole. However, for Dirac perturbation,  increasing the parameter $L$ prevents the rise of effective potential for both RNdS and RNAdS-like black holes.
We also investigate the greybody factor of  RNdS-like black hole only by using rigorous bound method. 
The greybody factor of Dirac field perturbation is discussed separately for the massless and massive cases.  The greybody factor of massless Dirac field perturbation  relies on the distance between cosmological and event horizons, and  decreases with increasing the distance between the horizons. For the massive Dirac particle, studying the expression of greybody factor bound analytically is tedious. The approximation method is used to analyse how the greybody factor bound relies on the mass of the Dirac particle.
 We see from  Eq. \eqref{eqn gbm} that the mass of the  Dirac particle tends to increase the argument  of $\sech$ function  and hence it lowers the greybody factor bound  thereby making the waves more difficult to transmit through the black hole.
It is noted that the behaviours of greybody factors are same for all the  perturbations. Increasing the parameter $L$ tends to increase the bound of greybody factors of the black holes in all perturbations which permit  more waves to reach a far distant observer but it has an opposite effect with the increase in $Q$.    Hence  the greybody factor which is similar to the transmission coefficient relies on the shape of the effective potential. If the potential is large, there is less probability for the scalar, electromagnetic and Dirac particles to transmit through the black hole, hence the greybody factor lower and vise-versa. 
Using 6th order WKB method and the Padé approximation, we numerically calculate the QNMs for the RNdS-like black hole with varying $Q$ and $L$. Both the real parts and magnitude  of imaginary parts of QNMs frequencies increase with the increase of $Q$ in all the perturbations which leads to increase oscillation frequency and damping rate. However it has an opposite effect  for increasing $L$ in all the perturbations of RNdS-like black hole. The results of the time-domain profile analysis for the evolution of scalar, electromagnetic and Dirac perturbations are  in good agreement with the quasinormal frequencies obtained by using the WKB 6th order and  Padé approximation. It is also noted that  the photon sphere radius is independent of cosmological constant $\Lambda$, but the shadow radius depends on $\Lambda$. The photon sphere radius decreases with the increase of $Q$ and $L$ for both RNdS and RNAdS-like black holes.  The shadow radius for RNdS-like black hole decreases with the increase in $Q$ and $L$. Further the radius of black hole shadow for RNdS-like black hole increases in the range $r_h<r_0\leq r_{max}$ and decreases in the range $r_{max}<r_0<r_c$.

Our analysis will contribute in expanding the theoretical framework of black hole physics by incorporating the Lorentz symmetry breaking arise from bumblebee field in the analysis of black hole perturbation. Furthermore, our investigation
of greybody factors and quasinormal modes along with the interpreting data from astrophysical observations such as  LIGO, Virgo can be used to further constrain the parameters of bumblebee gravity, offering a potential avenue to test Lorentz symmetry violation in the strong-field regime.

\appendix
\renewcommand{\theequation}{A\arabic{equation}}
\setcounter{equation}{0}

 \section*{Appendix A}
Putting the value of A(r) as given in the Eq. \eqref{eqn 2}, in Eq. \eqref{eqn 37} we can calculate effective potential for the massive case as follows
\begin{align}\label{eqn A1}
\int_{-\infty}^{\infty}|W^2|d{r_*}= \int_{r_h}^{r_c}\dfrac{2\omega(1+L) (\kappa^2+m^2r^2)^2}{r^2(2\omega\sqrt{1+L}(\kappa^2+m^2r^2)+(1-\frac{2M}{r}+\frac{2(1+L)}{2+L}\frac{Q^2}{r^2}-\frac{1}{3}(1+L)r^2\Lambda m \kappa}dr.
\end{align}
The integrand in the above equation can be factorised as 
\begin{align}
&\dfrac{m^4r^4+2\kappa^2m^2r^2+\kappa^4}{\left(2\omega \sqrt{1+L}m^2-\frac{1}{3}(1+L)\Lambda m \kappa\right)r^4+(2\omega\sqrt{1+L}\kappa^2+m\kappa)r^2-2mM\kappa r+\frac{2(1+L)Q^2m\kappa}{2+L}}\nonumber\\
&=\dfrac{m^4}{2\omega\sqrt{1+L}m^2-\frac{1}{3}(1+L)\Lambda m\kappa}
+\dfrac{ar^2+br+c}{D(r)},
\end{align}
where, 
\begin{align}
&a= 2k^2m^2- \dfrac{\kappa m^5+2k^2\sqrt{1+L}m^4\omega}{2\omega\sqrt{1+L}m^2-\frac{1}{3}(1+L)\Lambda m\kappa},\hspace{.5cm}
b= \dfrac{2\kappa m^5 M}{2\omega\sqrt{1+L}m^2-\frac{1}{3}(1+L)\Lambda m \kappa},\nonumber\\
&c= \kappa^4- \dfrac{2(1+L)Q^2\kappa m^5}{(2+L)\left(\omega\sqrt{1+L}m^2-\frac{1}{3}(1+L)\Lambda m \kappa\right)}.
\end{align}
We assume that $R_1, R_2, R_3$ and $ R_4$ are the roots of the equation
\begin{align}
&D(r)=\left(2\omega \sqrt{1+L}m^2-\frac{1}{3}(1+L)\Lambda m \kappa \right)r^4+(2\omega\sqrt{1+L}\kappa^2+m \kappa)r^2- 2mM \kappa r+\frac{2(1+L)Q^2m \kappa}{2+L}=0.
\end{align}
Applying partial fraction method,  we get
\begin{align}
\dfrac{ar^2+br+c}{(r-R_1)(r-R_2)(r-R_3)(r-R_4)}=\dfrac{Z_1}{(r-R_1)}+\dfrac{Z_2}{(r-R_2)}+\dfrac{Z_3}{(r-R_3)}+\dfrac{Z_4}{(r-R_4)}.
\end{align}
Putting $r=R_1$,  $r=R_2$,  $r=R_3$ and  $r=R_4$ in the above equation, the constant terms are derived as follows
\begin{align}
&Z_1=\dfrac{aR^2_1+bR_1+c}{(R_1-R_2)(R_1-R_3)(R_1-R_4)},\hspace{.5cm}
Z_2=\dfrac{aR^2_2+bR_2+c}{(R_2-R_1)(R_2-R_3)(R_2-R_4)},\nonumber\\
&Z_3=\dfrac{aR^2_3+bR_3+c}{(R_3-R_1)(R_3-R_2)(R_3-R_4)},\hspace{.5cm}
Z_4=\dfrac{aR^2_4+bR_4+c}{(R_4-R_1)(R_4-R_2)(R_4-R_3)}.
\end{align}
Then, we get
\begin{align}\label{eqn A7}
&\int_{r_h}^{r_c}\dfrac{ar^2+br+c}{(r-R_1)(r-R_2)(r-R_3)(r-R_4)}dr=Z_1\log\left|\frac{r_c-R_1}{r_h-R_1}\right|+Z_2\log\left|\frac{r_c-R_2}{r_h-R_2}\right|+Z_3\log\left| \frac{r_c-R_3}{r_h-R_3}\right|+Z_4\log\left| \frac{r_c-R_4}{r_h-R_4}\right|.
\end{align}
Using Eq. \ref{eqn A7} in Eq. \ref{eqn A1}, we get
\begin{align}
&\int_{-\infty}^{\infty}\left| W^2 \right | dr_*=2\omega(1+L)\left[ \dfrac{m^4}{2\omega\sqrt{1+L}m^2-\frac{1}{3}(1+L)\Lambda m \kappa }(r_c-r_h)+Z_1\log\left| \frac{r_c-R_1}{r_h-R_1}\right | \right. \nonumber\\ & \left.
+Z_2\log\left | \frac{r_c-R_2}{r_h-R_2}\right|
+Z_3\log \left | \frac{r_c-R_3}{r_h-R_3}\right | +Z_4\log\left|\frac{r_c-R_4}{r_h-R_4}\right | \right]. 
\end{align}
The greybody factor for the massive Dirac field perturbation which is based on rigorous bound technique is derived as
\begin{align}
&T\geq \sech^2 (1+L)\left[ \dfrac{m^4}{2\omega\sqrt{1+L}m^2-\frac{1}{3}(1+L)\Lambda m \kappa}(r_c-r_h)+Z_1\log\left| \frac{r_c-R_1}{r_h-R_1}\right |  \right. \nonumber\\ & \left.
+Z_2\log \left | \frac{r_c-R_2}{r_h-R_2} \right |
+Z_3\log \left| \frac{r_c-R_3}{r_h-R_3}\right| +Z_4\log \left|\frac{r_c-R_4}{r_h-R_4} \right | \right]. 
\end{align}


\end{document}